\documentclass[longauth]{aa}  
\usepackage{graphicx}
\usepackage{dblfloatfix}
\usepackage{lastpage}
\usepackage{lscape}
\usepackage{rotating}
\usepackage{tabularx}
\usepackage{xcolor}
\usepackage[normalem]{ulem}
\usepackage[]{footmisc}
\usepackage{orcidlink}
\usepackage{txfonts}
\usepackage{xspace}
\usepackage{gensymb}
\usepackage{hyperref}
\hypersetup{
    colorlinks=true,
    linkcolor=blue,
    filecolor=magenta,      
    urlcolor=blue,
    citecolor=blue,
}

\def\hess{H.E.S.S.\xspace}
\def\fermi{\textit{Fermi}-LAT\xspace}
\def\ssftt{SS~433\xspace}
\def\xray{X-ray\xspace}
\def\vsgr{V4641~Sgr\xspace}

\makeatletter
\def\blfootnote{\xdef\@thefnmark{}\@footnotetext}
\makeatother

\begin{document} 
\title{Constraining the nature of the most extreme Galactic particle accelerator. H.E.S.S. observations of the microquasar V4641 Sgr}	
\author {
{\small
A.~Acharyya\inst{\ref{USD}}\orcidlink{0000-0002-2028-9230} 
\and F.~Aharonian\inst{\ref{DIAS},\ref{MPIK}}\orcidlink{0000-0003-1157-3915} 
\and H.~Ashkar\inst{\ref{LLR}}\orcidlink{0000-0002-2153-1818} 
\and M.~Backes\inst{\ref{UNAM},\ref{NWU}}\orcidlink{0000-0002-9326-6400} 
\and R.~Batzofin\inst{\ref{UP}}\orcidlink{0000-0002-5797-3386} 
\and D.~Berge\inst{\ref{DESY},\ref{HUB}}\orcidlink{0000-0002-2918-1824} 
\and K.~Bernl\"ohr\inst{\ref{MPIK}}\orcidlink{0000-0001-8065-3252} 
\and M.~B\"ottcher\inst{\ref{NWU}}\orcidlink{0000-0002-8434-5692} 
\and C.~Boisson\inst{\ref{LUX}}\orcidlink{0000-0001-5893-1797} 
\and J.~Bolmont\inst{\ref{LPNHE}}\orcidlink{0000-0003-4739-8389} 
\and F.~Brun\inst{\ref{IRFU}}\orcidlink{0000-0003-0770-9007} 
\and B.~Bruno\inst{\ref{ECAP}}  
\and C.~Burger-Scheidlin\inst{\ref{DIAS}}\orcidlink{0000-0002-7239-2248} 
\and T.~Bylund\inst{\ref{LUX}}  
\and S.~Casanova\inst{\ref{IFJPAN}}\orcidlink{0000-0002-6144-9122} 
\and J.~Celic\inst{\ref{ECAP}}  
\and M.~Cerruti\inst{\ref{APC}}\orcidlink{0000-0001-7891-699X} 
\and A.~Chen\inst{\ref{Wits}}\orcidlink{0000-0001-6425-5692} 
\and M.~Chernyakova\inst{\ref{DCU},\ref{DIAS}}\orcidlink{0000-0002-9735-3608} 
\and J. O.~Chibueze\inst{\ref{NWU},\ref{UNAM}}\orcidlink{0000-0002-9875-7436} 
\and O.~Chibueze\inst{\ref{NWU}}\orcidlink{0000-0001-8601-2675} 
\and B.~Cornejo\inst{\ref{IRFU}}\orcidlink{0009-0003-0039-0483} 
\and G.~Cotter\inst{\ref{UOX}}\orcidlink{0000-0002-9975-1829} 
\and J.~de~Assis~Scarpin\inst{\ref{LLR}}\orcidlink{0009-0004-4411-236X} 
\and M.~de~Bony~de~Lavergne\inst{\ref{IRFU},\ref{CPPM}}\orcidlink{0000-0002-4650-1666} 
\and M.~de~Naurois\inst{\ref{LLR}}\orcidlink{0000-0002-7245-201X} 
\and E.~de~O\~na~Wilhelmi\inst{\ref{DESY}}\orcidlink{0000-0002-5401-0744} 
\and A.~G.~Delgado~Giler\inst{\ref{HUB}}\orcidlink{0000-0003-2190-9857} 
\and J.~Devin\inst{\ref{LUPM}}\orcidlink{0000-0003-1018-7246} 
\and A.~Djannati-Ata\"i\inst{\ref{APC}}\orcidlink{0000-0002-4924-1708} 
\and A.~Dmytriiev\inst{\ref{NWU}}\orcidlink{0000-0003-0102-5579} 
\and K.~Egberts\inst{\ref{UP}}\orcidlink{0009-0000-5511-7060} 
\and K.~Egg\inst{\ref{ECAP}}\orcidlink{0009-0002-4238-034X} 
\and J.-P.~Ernenwein\inst{\ref{CPPM}}  
\and C.~Esca\~{n}uela~Nieves\inst{\ref{MPIK}}\orcidlink{0000-0002-7297-8126} 
\and P.~Fauverge\inst{\ref{LP2I}}\orcidlink{0009-0006-1613-6633} 
\and K.~Feijen\inst{\ref{APC}}\orcidlink{0000-0003-1476-3714} 
\and M.~D.~Filipovic\inst{\ref{Sydney}}\orcidlink{0000-0002-4990-9288} 
\and G.~Fontaine\inst{\ref{LLR}}\orcidlink{0000-0002-6443-5025} 
\and S.~Funk\inst{\ref{ECAP}}\orcidlink{0000-0002-2012-0080} 
\and S.~Gabici\inst{\ref{APC}}  
\and Y.A.~Gallant\inst{\ref{LUPM}}  
\and J.F.~Glicenstein\inst{\ref{IRFU}}\orcidlink{0000-0003-2581-1742} 
\and J.~Glombitza\inst{\ref{ECAP}}\orcidlink{0000-0001-9683-4568} 
\and P.~Goswami\inst{\ref{LSW}}\orcidlink{0000-0001-5430-4374} 
\and M.-H.~Grondin\inst{\ref{LP2I}}\orcidlink{0000-0002-8383-251X} 
\and L.~Heckmann\inst{\ref{APC}}\orcidlink{0000-0002-6653-8407} 
\and B.~He{\ss}\inst{\ref{IAAT}}\orcidlink{0009-0004-9999-171X} 
\and J.A.~Hinton\inst{\ref{MPIK}}\orcidlink{0000-0002-1031-7760} 
\and W.~Hofmann\inst{\ref{MPIK}}\orcidlink{0000-0001-8295-0648} 
\and T.~L.~Holch\inst{\ref{DESY}}\orcidlink{0000-0001-5161-1168} 
\and M.~Holler\inst{\ref{Innsbruck}}\orcidlink{0000-0002-0107-8657} 
\and M.~Jamrozy\inst{\ref{OAUJ}}\orcidlink{0000-0002-0870-7778} 
\and F.~Jankowsky\inst{\ref{LSW}}  
\and A.~Jardin-Blicq\inst{\ref{LP2I}}\orcidlink{0000-0002-6738-9351} 
\and I.~Jaroschewski\inst{\ref{IRFU}}\orcidlink{0000-0001-5180-2845} 
\and D.~Jimeno\inst{\ref{DESY}}\orcidlink{0009-0001-2499-9467} 
\and I.~Jung-Richardt\inst{\ref{ECAP}}  
\and K.~Katarzy{\'n}ski\inst{\ref{NCUT}}\orcidlink{0000-0002-8806-4863} 
\and D.~Kerszberg\inst{\ref{LPNHE}}\orcidlink{0000-0002-5289-1509} 
\and B. Khélifi\inst{\ref{APC}}\orcidlink{0000-0001-6876-5577} 
\and N.~Komin\inst{\ref{LUPM},\ref{Wits}}\orcidlink{0000-0003-3280-0582} 
\and K.~Kosack\inst{\ref{IRFU}}\orcidlink{0000-0001-8424-3621} 
\and D.~Kostunin\inst{\ref{DESY}}\orcidlink{0000-0002-0487-0076} 
\and R.G.~Lang\inst{\ref{ECAP}}\orcidlink{0000-0003-0492-5628} 
\and S.~Lazarevi\'c\inst{\ref{Sydney}}\orcidlink{0000-0001-6109-8548} 
\and A.~Lemi\`ere\inst{\ref{APC}}  
\and M.~Lemoine-Goumard\inst{\ref{LP2I}}\orcidlink{0000-0002-4462-3686} 
\and J.-P.~Lenain\inst{\ref{LPNHE}}\orcidlink{0000-0001-7284-9220} 
\and P.~Liniewicz\inst{\ref{OAUJ}}\orcidlink{0009-0008-3575-3965} 
\and A.~Luashvili\inst{\ref{NWU}}\orcidlink{0000-0003-4384-1638} 
\and J.~Mackey\inst{\ref{DIAS}}\orcidlink{0000-0002-5449-6131} 
\and D.~Malyshev\inst{\ref{IAAT}}\orcidlink{0000-0001-9689-2194} 
\and V.~Marandon\inst{\ref{IRFU}}\orcidlink{0000-0001-9077-4058} 
\and M.~G.~F.~Mayer\inst{\ref{ECAP}}\orcidlink{0000-0002-9771-9841} 
\and A.~Mehta\inst{\ref{DESY}}  
\and A.M.W.~Mitchell\inst{\ref{ECAP}}\orcidlink{0000-0003-3631-5648} 
\and R.~Moderski\inst{\ref{NCAC}}\orcidlink{0000-0002-8663-3882} 
\and L.~Mohrmann\inst{\ref{MPIK}}\orcidlink{0000-0002-9667-8654} 
\and A.~Montanari\inst{\ref{LSW}}\orcidlink{0000-0002-3620-0173} 
\and E.~Moulin\inst{\ref{IRFU}}\orcidlink{0000-0003-4007-0145} 
\and J.~Niemiec\inst{\ref{IFJPAN}}\orcidlink{0000-0001-6036-8569} 
\and L.~Olivera-Nieto\inst{\ref{MPIK}}\hyperref[current]{{\large$^{,\star,}$}}{\large *}\orcidlink{0000-0002-9105-0518}
\and M.O.~Moghadam\inst{\ref{UP}}\orcidlink{0009-0003-2479-1863} 
\and S.~Panny\inst{\ref{Innsbruck}}\orcidlink{0000-0001-5770-3805} 
\and R.D.~Parsons\inst{\ref{HUB}}\orcidlink{0000-0003-3457-9308} 
\and U.~Pensec\inst{\ref{LPNHE}}\orcidlink{0009-0009-2359-1775} 
\and P.~Pichard\inst{\ref{APC}}\orcidlink{0009-0005-9803-0762} 
\and T.~Preis\inst{\ref{Innsbruck}}\orcidlink{0009-0001-7110-6764} 
\and G.~P\"uhlhofer\inst{\ref{IAAT}}\orcidlink{0000-0003-4632-4644} 
\and M.~Punch\inst{\ref{APC}}\orcidlink{0000-0002-4710-2165} 
\and A.~Quirrenbach\inst{\ref{LSW}}  
\and A.~Reimer\inst{\ref{Innsbruck}}\orcidlink{0000-0001-8604-7077} 
\and O.~Reimer\inst{\ref{Innsbruck}}\orcidlink{0000-0001-6953-1385} 
\and I.~Reis\inst{\ref{IRFU}}  
\and Q.~Remy\inst{\ref{MPIK}}$^{,}${\large *} \orcidlink{0000-0002-8815-6530}
\and H.~X.~Ren\inst{\ref{MPIK}}\orcidlink{0000-0003-0221-2560} 
\and B.~Reville\inst{\ref{MPIK}}$^{,}${\large *} \orcidlink{0000-0002-3778-1432}
\and F.~Rieger\inst{\ref{MPIK}}  
\and G.~Roellinghoff\inst{\ref{ECAP}}\orcidlink{0000-0002-9824-9597} 
\and G.~Rowell\inst{\ref{Adelaide}}\orcidlink{0000-0002-9516-1581} 
\and B.~Rudak\inst{\ref{NCAC}}\orcidlink{0000-0003-0452-3805} 
\and K.~Sabri\inst{\ref{LUPM}}  
\and S.~Safi-Harb\inst{\ref{SS}}\orcidlink{0000-0001-6189-7665}
\and V.~Sahakian\inst{\ref{YPI}}\orcidlink{0000-0003-1198-0043} 
\and A.~Santangelo\inst{\ref{IAAT}}\orcidlink{0000-0003-4187-9560} 
\and M.~Sasaki\inst{\ref{ECAP}}\orcidlink{0000-0001-5302-1866} 
\and F.~Sch\"ussler\inst{\ref{IRFU}}\orcidlink{0000-0003-1500-6571} 
\and J.N.S.~Shapopi\inst{\ref{UNAM}}\orcidlink{0000-0002-7130-9270} 
\and W.~Si~Said\inst{\ref{LLR}}\orcidlink{0009-0007-6555-6893} 
\and H.~Sol\inst{\ref{LUX}}  
\and {\L.}~Stawarz\inst{\ref{OAUJ}}\orcidlink{0000-0002-7263-7540} 
\and S.~Steinmassl\inst{\ref{MPIK}}\orcidlink{0000-0002-2865-8563} 
\and T.~Tanaka\inst{\ref{Konan}}\orcidlink{0000-0002-4383-0368} 
\and A.M.~Taylor\inst{\ref{DESY}}\orcidlink{0000-0001-9473-4758} 
\and G.~L.~Taylor\inst{\ref{LSW}}\orcidlink{0009-0001-8062-036X} 
\and R.~Terrier\inst{\ref{APC}}\orcidlink{0000-0002-8219-4667} 
\and Y.~Tian\inst{\ref{DESY}}\orcidlink{0009-0005-7165-3791} 
\and A.~Timmermans\inst{\ref{MPIK}}\orcidlink{0009-0000-8234-9828} 
\and M.~Tsirou\inst{\ref{DESY}}\orcidlink{0000-0003-3417-1425} 
\and N.~Tsuji\inst{\ref{ICRR}}$^{,}${\large*}\orcidlink{0000-0001-7209-9204}
\and T.~Unbehaun\inst{\ref{ECAP}}\orcidlink{0000-0002-7378-4024} 
\and C.~van~Eldik\inst{\ref{ECAP}}\orcidlink{0000-0001-9669-645X} 
\and M.~Vecchi\inst{\ref{Groningen}}\orcidlink{0000-0002-5338-6029} 
\and C.~Venter\inst{\ref{NWU}}\orcidlink{0000-0002-2666-4812} 
\and J.~Vink\inst{\ref{GRAPPA}}\orcidlink{0000-0002-4708-4219} 
\and V.~Voitsekhovskyi\inst{\ref{GRAPPA}}\orcidlink{0000-0002-3906-4840} 
\and S.J.~Wagner\inst{\ref{LSW}}\orcidlink{0000-0002-7474-6062} 
\and A.~Wierzcholska\inst{\ref{IFJPAN},\ref{LSW}}\orcidlink{0000-0003-4472-7204} 
\and M.~Zacharias\inst{\ref{LSW},\ref{NWU}}\orcidlink{0000-0001-5801-3945} 
\and A.A.~Zdziarski\inst{\ref{NCAC}}\orcidlink{0000-0002-0333-2452} 
\and A.~Zech\inst{\ref{LUX}}  
\and W.~Zhong\inst{\ref{DESY}}\orcidlink{0000-0003-3717-2861} 
\\ (H.E.S.S.\ Collaboration)
\and \\ S.~Takekawa\inst{\ref{36}}\orcidlink{0000-0001-8147-6817} 
}}

\institute{
University of Southern Denmark \label{USD}
\and Astronomy $\&$ Astrophysics Section, School of Cosmic Physics, Dublin Institute for Advanced Studies, DIAS Dunsink Observatory, Dublin D15 XR2R, Ireland \label{DIAS}
\and Max-Planck-Institut für Kernphysik, P.O. Box 103980, D 69029 Heidelberg, Germany \label{MPIK}
\and Laboratoire Leprince-Ringuet, École Polytechnique, CNRS, Institut Polytechnique de Paris, F-91128 Palaiseau, France \label{LLR}
\and University of Namibia, Department of Physics, Private Bag 13301, Windhoek 10005, Namibia \label{UNAM}
\and Centre for Space Research, North-West University, Potchefstroom 2520, South Africa \label{NWU}
\and Institut für Physik und Astronomie, Universität Potsdam, Karl-Liebknecht-Strasse 24/25, D 14476 Potsdam, Germany \label{UP}
\and Deutsches Elektronen-Synchrotron DESY, Platanenallee 6, 15738 Zeuthen, Germany \label{DESY}
\and Institut für Physik, Humboldt-Universität zu Berlin, Newtonstr. 15, D 12489 Berlin, Germany \label{HUB}
\and LUX, Observatoire de Paris, Université PSL, CNRS, Sorbonne Université, 5 Pl. Jules Janssen, 92190 Meudon, France \label{LUX}
\and Sorbonne Université, CNRS/IN2P3, Laboratoire de Physique Nucléaire, et de Hautes Energies, LPNHE, 4 place Jussieu, 75005 Paris, France \label{LPNHE}
\and IRFU, CEA, Université Paris-Saclay, F-91191 Gif-sur-Yvette, France \label{IRFU}
\and Friedrich-Alexander-Universität Erlangen-Nürnberg, Erlangen Centre for Astroparticle Physics,  Nikolaus-Fiebiger-Str. 2, 91058 Erlangen, Germany \label{ECAP}
\and Instytut Fizyki Jac{a}drowej PAN, ul. Radzikowskiego 152, ul. Radzikowskiego 152, 31-342 Kraków, Poland \label{IFJPAN}
\and Université Paris Cité, CNRS, Astroparticule et Cosmologie, F-75013 Paris, France \label{APC}
\and School of Physics, University of the Witwatersrand, 1 Jan Smuts Avenue, Braamfontein, Johannesburg, 2050, South Africa \label{Wits}
\and School of Physical Sciences and Centre for Astrophysics $\&$ Relativity, Dublin City University, Glasnevin, Dublin D09 W6Y4, Ireland \label{DCU}
\and University of Oxford, Department of Physics, Denys Wilkinson Building, Keble Road, Oxford OX1 3RH, UK, United Kingdom \label{UOX}
\and Aix Marseille Université, CNRS/IN2P3, CPPM, Marseille, France \label{CPPM}
\and Laboratoire Univers et Particules de Montpellier, Université Montpellier, CNRS/IN2P3, CC 72, Place Eugène Bataillon, F-34095 Montpellier Cedex 5, France \label{LUPM}
\and Université Bordeaux, CNRS, LP2I Bordeaux, UMR 5797, F-33170 Gradignan, France \label{LP2I}
\and School of Science, Western Sydney University, Locked Bag 1797, Penrith South DC, NSW 2751, Australia \label{Sydney}
\and Landessternwarte, Universität Heidelberg, Königstuhl, D 69117 Heidelberg, Germany \label{LSW}
\and Institut für Astronomie und Astrophysik, Universität Tübingen, Sand 1, D 72076 Tübingen, Germany \label{IAAT}
\and Universität Innsbruck, Institut für Astro- und Teilchenphysik, Technikerstraße 25, 6020 Innsbruck, Austria \label{Innsbruck}
\and Obserwatorium Astronomiczne, Uniwersytet Jagielloński, ul. Orla 171, 30-244 Kraków, Poland \label{OAUJ}
\and Institute of Astronomy, Faculty of Physics, Astronomy and Informatics, Nicolaus Copernicus University, Grudziadzka 5, 87-100 Torun, Poland \label{NCUT}
\and Nicolaus Copernicus Astronomical Center, Polish Academy of Sciences, ul. Bartycka 18, 00-716 Warsaw, Poland \label{NCAC}
\and GRAPPA, Anton Pannekoek Institute for Astronomy, University of Amsterdam, Science Park 904, 1098 XH Amsterdam, The Netherlands \label{GRAPPA}
\and Department of Physics and Astronomy, University of Manitoba, Winnipeg, MB, R3T 2N2, Canada\label{SS}
\and School of Physical Sciences, University of Adelaide, Adelaide 5005, Australia \label{Adelaide}
\and Yerevan Physics Institute, 2 Alikhanian Brothers St., 0036 Yerevan, Armenia \label{YPI}
\and Department of Physics, Konan University, 8-9-1 Okamoto, Higashinada, Kobe, Hyogo 658-8501, Japan \label{Konan}
\and Institute for Cosmic Ray Research, University of Tokyo, 5-1-5, Kashiwa-no-ha, Kashiwa, Chiba 277-8582, Japan \label{ICRR}
\and Kapteyn Astronomical Institute, University of Groningen, Landleven 12, 9747 AD Groningen, The Netherlands \label{Groningen}
\and Department of Applied Physics, Faculty of Engineering, Kanagawa University, 3-27-1 Rokkakubashi, Kanagawa-ku, Yokohama, Kanagawa 221-8686, Japan\label{36}}

\date{Received 3 October 2025 / Accepted 6 November 2025}
\offprints{
\protect\\\email{\href{mailto:contact.hess@hess-experiment.eu}{contact.hess@hess-experiment.eu};}
\protect\\\protect * Corresponding authors}
\abstract
{Microquasars have emerged as promising candidates to explain the cosmic-ray flux at petaelectronvolt energies. LHAASO observations revealed \vsgr as the most extreme example so far. Its gamma-ray spectrum extends up to 800~TeV, which requires particles with multi-PeV energy. The TeV emission is highly extended, which challenges expectations given the reported low-inclination angle of the \vsgr jets.
}
{We spatially and spectrally resolved the gamma-ray emission from \vsgr and investigated the particle acceleration in the system. }
{Using $\approx$100~h of \hess data, we performed a spectro-morphological study of the gamma-ray emission around \vsgr. We employed HI and dedicated CO observations of the region to infer the target material for cosmic-ray interactions.}
{We detected multi-TeV emission around \vsgr with a high significance. The emission region is elongated, and its major and minor axes are ${0.34\degree}^{\pm 0.04_{stat}}_{\pm 0.01_{syst}}$ and ${0.06\degree}^{\pm 0.01_{stat}}_{\pm 0.01_{syst}}$, respectively. We found a power-law spectrum with an index $\approx$1.8, and together with results from other gamma-ray instruments, this reveals a spectral energy distribution (SED) that peaks at energies of $\approx$100~TeV for the first time. 
We found indications (3$\sigma$) of a two-component morphology, with indistinguishable spectral properties. The position of \vsgr is inconsistent with the best-fit position of the single-component model and with the dip between the two components. We found no significant evidence of an energy-dependent morphology. No dense gas was found at any distance towards \vsgr, which places an upper limit of n$_{\rm gas}\lesssim 0.2$~cm$^{-3}$ within the gamma-ray emission region.}
{The peak of the SED at $\approx$100~TeV identifies \vsgr as a candidate cosmic-ray accelerator beyond the so-called knee. The absence of dense target gas places stringent energetic constraints on hadronic interpretations, however. The \hess measurement requires an unusually hard ($\approx 1.5$) spectral index for the protons. A leptonic scenario faces fewer obstacles if the particle transport is fast enough to avoid losses and to reproduce the observed energy-independent morphology. The absence of bright \xray emission across the gamma-ray emission region requires a magnetic field strength $\lesssim3$~$\mu$G, however. Our findings favour a leptonic origin of the gamma-ray emission. This conclusion does not exclude hadron acceleration in the \vsgr system.}


\authorrunning{H.E.S.S. Collaboration}
\titlerunning{H.E.S.S. observations of the microquasar V4641 Sgr}

\maketitle


\section{Introduction}

The origin of the cosmic-ray spectrum measured at Earth remains one of the largest mysteries in high-energy astrophysics. 
Precise measurements have revealed several spectral features that were interpreted as a result of differences in the production processes or astrophysical source classes involved, and also cosmic-ray propagation through space. At energies of a few petaelectronvolts, a spectral break (usually referred to as the knee) was measured~\citep{Amenomori2008,Aartsen2019, Collaboration2025}. Below this energy, convincing theoretical arguments and experimental evidence have suggested that particles are predominantly accelerated in supernova remnants ~\cite[SNRs, see e.g.][]{Ginzburg, Hillas2005,Bell2014}. Because cosmic rays arriving at Earth have undergone multiple scatterings in the turbulent magnetic field of the Galaxy, astrophysical sources of high-energy particles are better studied through neutral secondaries produced via interactions of particles with their source environment: photons and neutrinos. Gamma-ray observations in the TeV domain have identified hundreds of sources with a wide variety of types, including SNRs. The observed gamma-ray spectra of SNRs show cut-offs in energy at a few dozen TeV, however, corresponding to sub-PeV primary particles~\cite[e.g.][]{HESSCollaboration2018,2020ApJ...894...51A}. Even higher gamma-ray energies have only become accessible in recent years, but observations have yet to reveal an SNR source whose emission extends to the required energies to account for the knee feature~\citep{LHAASO2024}.

As a consequence, other source classes have increasingly been considered as accelerators of the highest-energy Galactic cosmic rays. One such class of objects are microquasars. These are stellar binary systems hosting a neutron star or a black hole (\xray binaries) and launch powerful and often relativistic jets. Microquasar jets have long been proposed as sites for acceleration to TeV and PeV energies~\cite[e.g.][]{ Aharonian2004,BoschRamon2005a,Cooper2020,Kantzas2023}, and experimental evidence supporting these claims has increased recently. In particular, the first report of TeV emission from a microquasar (\ssftt) by the High-Altitude Water Cherenkov (HAWC) Observatory \citep{ss433hawc} and the follow-up high-resolution study with the High Energy Stereoscopic System ~\citep[\hess,~][]{ss433} revealed efficient particle acceleration sites in the jets at a distance of some dozen parsecs~(pc) from the central binary. A second TeV detection of a microquasar (\vsgr) was reported by the HAWC Collaboration~\citep{hawcv4641}, which prompted follow-up observations with \hess The result of these observations is reported in this paper. The Large High Altitude Air Shower Observatory (LHAASO) recently reported the detection of gamma-rays with energies $>25$~TeV from these two systems and three additional accreting black-hole binary systems~\citep{lhaaso}. In particular, the LHAASO spectrum of \vsgr extends to 800~TeV at least, which means that particle acceleration to multi-PeV energies is required in the system. 

Among other PeV accelerator candidates, \vsgr stands out as an ideal laboratory for the study of cosmic-ray acceleration. Unlike many of the sources detected by LHAASO above some hundred TeV, \vsgr lies several degrees off the Galactic plane and not in a crowded star-forming region~\citep{LHAASOCygnus}. Accordingly, the identification of the microquasar as the counterpart of the gamma-ray source is firm. Additionally, protons and other nuclei are known to be present in microquasar jets~\citep{Margon1979a, DiazTrigo2013}. Specifically, emission lines from iron and other metals have been attributed to the jets of \vsgr~\citep{Zand2000}.
Regardless of whether the observed ultra-high-energy photons originate from inverse-Compton (IC) scattering of PeV electrons, or from neutral-pion decay following inelastic collisions of multi-PeV nuclei with ambient gas, the idea that these sources contribute to the local cosmic-ray spectrum above PeV energies remains compelling.

We present the \hess view of \vsgr. \hess is an array of imaging atmospheric Cherenkov telescopes (IACTs) that provides a superior angular and spectral resolution compared to the HAWC and LHAASO arrays of water Cherenkov detectors (WCDs)~\citep[e.g.][]{Hinton2009}. This enables a detailed study of the gamma-ray emission properties of \vsgr.

The paper is organised as follows. In Section~\ref{sec:the_guy} we introduce the microquasar \vsgr and summarise the knowledge about the system. We then describe the \hess telescopes and the procedure we used to select and analyse the data in Section~\ref{sec:the_data}. We report the results of the spectral and morphological characterisation of the emission in Section~\ref{sec:the_results} and on the dedicated radio observations in Section~\ref{sec:the_gas_main}. These are combined with insights from previous multi-wavelength observations to discuss the physical implications in Section~\ref{sec:the_discussion}. Finally, our conclusions are collected in Section~\ref{sec:the_conclusions}.

\section{The microquasar \vsgr}
\label{sec:the_guy}

\vsgr (SAX~J1819.3$-$2525, XTE~J1819$-$254), located at Galactic coordinates $l$=6.7740$\degree$ and  $b$=-4.7891$\degree$, is a low-mass Galactic \xray binary (LMXB) system with a period of 2.81728 days~\citep{Goranskij2001, Orosz2001}. It was identified as such in 1999, when it underwent a strong outburst~\citep{Markwardt1999,Zand2000, Wijnands2000} during which its \xray luminosity increased by several orders of magnitude. Modelling of the binary orbit yielded an estimate for the mass of the compact object in \vsgr of $\mathrm{M}_{\mathrm{BH}}\approx$7~M$_{\odot}$~\citep{Goranskij2003, Orosz2001, MacDonald2014}, which firmly identified it as a black hole. The companion star mass is estimated to be $\mathrm{M}_{*}\approx$3~M$_{\odot}$, making it one of the most massive companions observed in a LMXB system. The mass of the companion constrains the system age to $<1$~Gyr~\citep{Salvesen2020}.

During the \xray outburst in 1999, radio observations revealed an extended, fast-changing radio source typical of relativistic outflows~\citep{Hjellming2000}. The proper motions of the radio source were claimed to be (apparently) super-luminal~\citep{Hjellming2000}, which has been interpreted as evidence for the existence of a relativistic jet propagating with a low inclination angle with respect to our line of sight~\citep{Orosz2001, Salvesen2020}. However,~\citet{Hjellming2000} did not directly observe fast proper motions: they were inferred under the assumption that the radio blobs were launched at the same time as the \xray outburst.

Additional \xray observations of the same outburst revealed the presence of bright, relativistic iron lines~\citep{Zand2000, Miller2002} in addition to an unusually broad emission feature in the \xray spectrum above 4~keV. This feature has been interpreted as a blend of Doppler-shifted Fe lines, much like in the jets of \ssftt~\citep{Marshall2002}, or alternatively similar to the spectra of blazars at the onset of an inverse Compton hump~\citep{Gallo2014}. Both interpretations require the axis of the jets to be closely aligned with our line of sight, consistent with the interpretation of the radio observations. Optical~\citep{Revnivtsev2002a} and \xray~\citep{Revnivtsev2002} observations during the 1999 outburst suggest the presence of an extended envelope obscuring the source, which required mass accretion rates during the outburst exceeding the Eddington limit of $\dot{M} c^2>$ L$_{\mathrm{Edd}} = 1.26\times 10^{38}\left(\mathrm{M}_{\mathrm{BH}}/M_\odot\right) $~erg~s$^{-1}\approx 10^{39}$~erg~s$^{-1}$.

While the claim of super-luminal motion would suggest a jetted source that points towards us, studies of the inclination of the orbital plane provide contradictory results, or at least imply that the jet is not perpendicular to the orbital plane. Modelling of the binary orbit suggests that the orbital plane has an inclination $i_{\mathrm{orb}}=72.3\pm4.1\degree$~\citep{Orosz2001, MacDonald2014}, which has been cited as evidence of precession~\cite[e.g.][]{Gallo2014}. Such a large misalignment with the jet inclination inferred from the observation of apparent super-luminal motions ($i_{\mathrm{jet}}<$16$\degree$) cannot be explained via a natal kick from the progenitor supernova event~\citep{Salvesen2020}. Additionally, estimates of the inclination of the inner accretion disk yield a relatively high value of $i_{\mathrm{disk}}=43\pm15\degree$~\citep{Miller2002}. \cite{Martin2008} considered the possibility that these mismatches could be explained by a warped accretion disk, though concluded that this is not a viable explanation and that the observed $i_{\mathrm{disk}}$ and $i_{\mathrm{jet}}$ are in tension. Unfortunately, despite subsequent less powerful radio and \xray flares~\cite[e.g.][]{MunozDarias2018,Shaw2022}, there have been no further direct detections of the jet since the 1999 outburst, and thus uncertainty about the geometry of the system remains.

The constraints on the distance of the binary system are less ambiguous.
While~\cite{Hjellming2000} assumed a modest distance of 0.5~kpc, the earliest quantitative estimate of the distance to \vsgr was made by \cite{Orosz2001} by determining the spectral type of the companion star during \xray quiescence and comparing it to the observed magnitude accounting for extinction, finding $d=9.6\pm2.5$~kpc. This calculation was updated by~\cite{MacDonald2014}, which accounted for optical variability during the \xray quiescence period and yielded $d=6.2\pm0.7$~kpc. Measurements of parallax with the Gaia satellite~\citep{BailerJones2018} provide an independent distance estimate. A dedicated analysis of the Gaia data, taking into account the proper motion of the binary~\citep{Ghandi2018}, yields a consistent estimate of $d=6.62\pm1.81$~kpc. For simplicity we will adopt $d=6.2$~kpc throughout this work, unless specified otherwise. The distance could however plausibly lie somewhere between 5 and 8~kpc.

The identification of \vsgr as a gamma-ray source was reported by the HAWC collaboration, who detected an extended source around the microquasar~\citep{hawcv4641}. Due to its location below the Galactic plane, it is unlikely that another unknown Galactic counterpart could account for the observed gamma rays. Analysis of the HAWC source revealed that the best-fit morphological model is one extended ($\sigma_{\mathrm{major}}=0.54\pm0.08\pm0.1\degree$, which de-projected corresponds to 58.4$d_{\rm 6.2kpc}$~pc) and highly elliptical ($e=0.98\pm0.01$) component with a power-law spectrum of index $2.1\pm0.1$~\citep{hawcv4641}. A model with two point sources at distances of 0.23$\degree$ and 0.46$\degree$ from the binary was also considered, but could not be statistically distinguished from the single extended model. Measurements by LHAASO reveal a source with consistent morphology whose spectrum reaches up to at least 800~TeV~\citep{lhaaso}, firmly ruling out an extragalactic origin. In the GeV band, \cite{Zhao2025} analysed 15~years of data from the \textit{Fermi} Large Area Telescope (LAT) and reported no detected emission across the entire region, neither steady nor variable. They derive upper limits using a spatial template constructed with the LHAASO significance map, which suffers from being convolved with the LHAASO angular resolution of tens of arcmin. 

It is not understood why a jet of reported low angle with our line of sight would produce such extended and asymmetric gamma-ray emission, measuring over 100~pc across when projected onto the sky. One possible explanation could be the proper motion of the source. However, the direction of the proper motion of \vsgr does not align with the orientation of the observed gamma-ray emission, but is instead orientated directly upwards towards the Galactic plane~\cite[see Figure 3 in][]{Ghandi2018}. The proper motion can also be used to estimate the time that \vsgr has spent at its current position. Adopting a distance of 6.2~kpc places \vsgr more than 500~pc below the Galactic plane (400 and 670~pc for 5 and 8~kpc, respectively). \xray binary systems located far from the Galactic Plane are usually explained as a consequence of natal kicks from their progenitor supernova event~\citep[e.g.][]{Blaauw1961}. \cite{Salvesen2020} traced the orbital trajectory of \vsgr backwards in time through the Milky Way galaxy for 1~Gyr, using its current position and velocity as initial conditions. The authors constrain the most recent Galactic-plane crossing to have happened approximately 10~Myr ago. This implies that \vsgr has moved at least $\approx5\degree$ in Galactic latitude in that time, implying that $\approx 2$~Myr ago, \vsgr was likely located $\approx 1\degree$ away from its current location. Consequently, even if the system were older than a few million years, any earlier evidence of particle acceleration would not be found at its current location.

The detection by HAWC prompted a renewed interest in \vsgr at different wavelengths. X-Ray Imaging and Spectroscopy Mission (XRISM) observations of the source, following an \xray flare in September 2024, revealed the presence of an additional extended \xray emission around the binary~\citep{Suzuki2025}. The reported \xray emission has a much smaller spatial extent than the extended HAWC detection, with a Gaussian radius of $7\pm3$~arcmin (the field-of-view of the Xtend instrument aboard XRISM is 38$\times$38~arcmin), but large enough to be independent of the flaring episode since the corresponding physical extent is $(13\pm5$)$d_{\rm 6.2kpc}$~pc. The nature of the emission (thermal/non-thermal) could not be determined. If it is of thermal origin, it would require jet luminosities comparable to L$_{\mathrm{Edd}}$. If instead, the observed emission is synchrotron radiation from a completely unrelated population of accelerated electrons, it would require $B \gtrsim 8~\mu$G so as to not over-produce gamma~rays via inverse Compton scattering in the XRISM emission region compared to the spectra measured by HAWC~\citep{Suzuki2025}.

\section{Observations and data analysis}
\label{sec:the_data}

\subsection{The \hess telescopes}
\label{subsec:hess}
\hess is an array of IACTs located in Namibia, sensitive to gamma~rays ranging from tens of GeV to tens of TeV. The array consists of five Cherenkov telescopes: four with mirror diameters of 12 m (referred to as CT1–CT4) placed in a square configuration and a single telescope at the centre (CT5) with a mirror diameter of 28 m. The 12-m telescope array is sensitive to gamma~rays of energies above $\approx$100~GeV~\citep{Aharonian2006} and has a field-of-view (FoV) of 5$\degree$ diameter, while the central, large telescope is able to detect fainter Cherenkov light, which in turn translates to lower gamma-ray energies~\citep{Holler2015, Unbehaun2025}. \hess started operations in 2003, when it comprised only CT1-4. This period of operation is referred to as the \hess I phase. The cameras of these telescopes were upgraded in 2016~\citep{Ashton2020}, starting the phase referred to as \hess IU. The central telescope, CT5, was built in 2012, and the full array configuration is referred to as \hess II. The CT5 camera was upgraded in 2019~\citep{Bi2022}. 

\subsection{Data selection and event reconstruction}
\label{subsec:selection}
\hess observations are conducted in runs of typical duration 28~min. The region around \vsgr was observed for a total of only 16 runs in the \hess I era~\citep{HESSCollaboration2018a}. The bulk of the data was taken following the report by the HAWC observatory in 2022\footnote{J. Goodman, Contribution to the 7th Heidelberg International Symposium on High-Energy Gamma-Ray Astronomy} using the \hess IU cameras.  The system was observed in 2022, 2023 and 2024, resulting in a total of 268 runs, corresponding to almost 125~h of observations. Criteria to select good-quality runs include atmospheric transparency (ranging between 0.7 and 1.3, where 1 is the nominal value) and night-sky-background noise (NSB). When runs were taken with normal trigger settings, an NSB threshold of 700~MHz (measured by CT5) was applied to remove runs taken under bright moonlight conditions. This NSB noise cut is not required for runs taken with specific moonlight trigger settings. After these quality cuts, a total of 211 runs remained, 84 of which taken using specific moonlight trigger settings, which translates to a total observing time slightly above 100~h. Correcting for the fact that the acceptance is not uniform across the FoV and instead decreases with increasing offset from the pointing position, the equivalent on-axis exposure is 97~h. Because the majority of the data is the result of dedicated observations with pointing positions distributed around \vsgr with offsets of 0.7$\degree$, the exposure across the region of interest is uniform, with less than 1\% variation within a radius of 0.7$\degree$.

\hess records stereoscopic images of atmospheric showers created by high-energy particles or gamma~rays as they impact the Earth’s atmosphere. After an initial reconstruction of shower geometry~\citep{Hillas1985} the shower images were selected for gamma-likeness using a Boosted Decision Tree classifier as described in~\cite{Ohm2009}. The classifier cuts are optimised for a faint source with a $\frac{dN}{dE}\propto E^{-2}$ spectral shape, usually referred to as hard cuts. These cuts include a 200~photo-electron requirement for the total charge deposited by the shower in at least one of the images recorded. This requirement increases the energy threshold, but also improves the signal-to-noise ratio, especially at several TeV. Final event parameters were determined by reconstructing the selected events using the ImPACT~\citep{Parsons2014} algorithm, which uses a maximum-likelihood framework to fit a library of simulated templates to the recorded images. This process resulted in an estimate for the gamma-ray energy and direction, among other parameters. Using ImPACT for the event reconstruction translates to significant improvements in the angular resolution, which for this analysis resulted in a 68\% containment of the point-spread function (PSF) which ranges between 0.05$\degree$ (below 10~TeV) and 0.06$\degree$ (above 10~TeV). In the data analysis presented in this paper, only information from CT1-4 was used for these steps. 

The instrument response functions (IRFs) were derived from Monte Carlo simulations of air showers which were then reconstructed in the same way as the data. A background model, which describes the properties of the hadronic background that is incorrectly classified as gamma-like, was derived using observation runs with mostly extragalactic pointing positions in which expected gamma-ray sources are masked. The background model varies between runs due to different pointing altitude positions and hardware epochs~\citep{Mohrmann2019}.

\subsection{Data reduction and analysis}
\label{subsec:reduction}
We used \textit{Gammapy}-1.2~\citep{Donath2023, GammapyZenodo} to reduce and analyse the data. The selected gamma-like events were binned into a three-dimensional square sky-map of 6$\degree$ width and 0.01$\degree$ bin size, centred at the position of \vsgr. The third dimension corresponds to the energy axis, which comprises 16
bins equally spaced in logarithmic energy between 0.4 and 80~TeV. For each run, a safe energy range was derived, defined as the range in which the energy is well-reconstructed, i.e. the energy bias is less than 10\%~\citep{Aharonian2006}, resulting in an energy range of 0.8 to 22~TeV across the dataset.

For each individual run, the counts predicted by the background model outside of a source exclusion mask were fit to those measured in the same region using two parameters that modify the overall background normalisation and spectral shape, respectively. This procedure corrects the background model for possible variations due to atmospheric conditions and instrumental characteristics, as described in~\cite{Mohrmann2019}. The exclusion mask defined to cover known and expected gamma-ray sources in the \vsgr field of view is composed of a circle centred on the position of \vsgr with radius 0.8$\degree$, a circle of 0.2$\degree$ radius around the position of M28, a nearby ($1.3\degree$ from \vsgr) globular cluster known to host several millisecond pulsars, and a circle of radius 0.5$\degree$ around the closest ($2.8\degree$ from \vsgr) known GeV pulsar, PSR~J1809-2332~\citep{Abdo2010}.

The IRFs were projected onto multi-dimensional sky-maps. The set of maps corresponding to each observation was stacked by adding the counts and background maps and combining the IRFs weighted by the exposure of each run, considering only the energy range determined as safe for each observation. The analysis results were confirmed by an independent analysis chain. This cross-check analysis employs an independent calibration, reconstruction and background suppression~\citep{Naurois2009}.

Significance maps were generated using a maximum-likelihood ratio test which compared the number of measured counts to the expected counts from the combination of a source and the background. In particular, we produced significance maps assuming a spectral index of 2, and two different spatial assumptions: a Gaussian kernel of full-width at half maximum 0.06$\degree$ (similar to the 68\% containment of the \hess PSF) and another one of 0.12$\degree$ to facilitate visual comparisons with the HAWC significance maps. The final significance map reflects smoothing by the combination of these kernels with the \hess PSF.

To describe the properties of the emission, we fitted combined spatial and spectral models within a maximum-likelihood framework using \textit{Gammapy}. For each model, we derived the test-statistic (TS) as the Poisson log-likelihood of the data. The difference in TS between two nested models follows a $\chi^2$ distribution with N$_{\rm dof}$ degrees of freedom, where N$_{\rm dof}$ is the difference in number of parameters between the nested models. We used this relation to convert $\Delta$TS to significances ($\sigma$). 

For each model component, we derived flux points by splitting the entire energy range into smaller bins and fitting a normalisation factor to the prediction of the model in each bin, keeping the spectral shape fixed to that of the global fit. If the significance of the model in that energy bin is less than 2$\sigma$, the 95\% confidence upper limit was derived and shown instead. 

In addition to flux maps, we derived flux profiles along a given spatial direction by defining rectangular regions orthogonal to that direction. Inside of each region, the measured counts were compared to the background and fitted to a spectral model with fixed spectral index. We compared this approach to using rectangular spatial templates in a spectro-morphological model, which would take into account contamination between regions due to the finite PSF. This approach yielded consistent results. We therefore kept the simple approach to avoid the need for spatial assumptions within the rectangular regions.

Systematic uncertainties in the model parameters were calculated with a Monte Carlo-based approach following~\cite{Collaboration2023}. Details are provided in Appendix~\ref{app:systematics}.

\begin{figure}
	\centering
		\includegraphics[width=.95\linewidth]{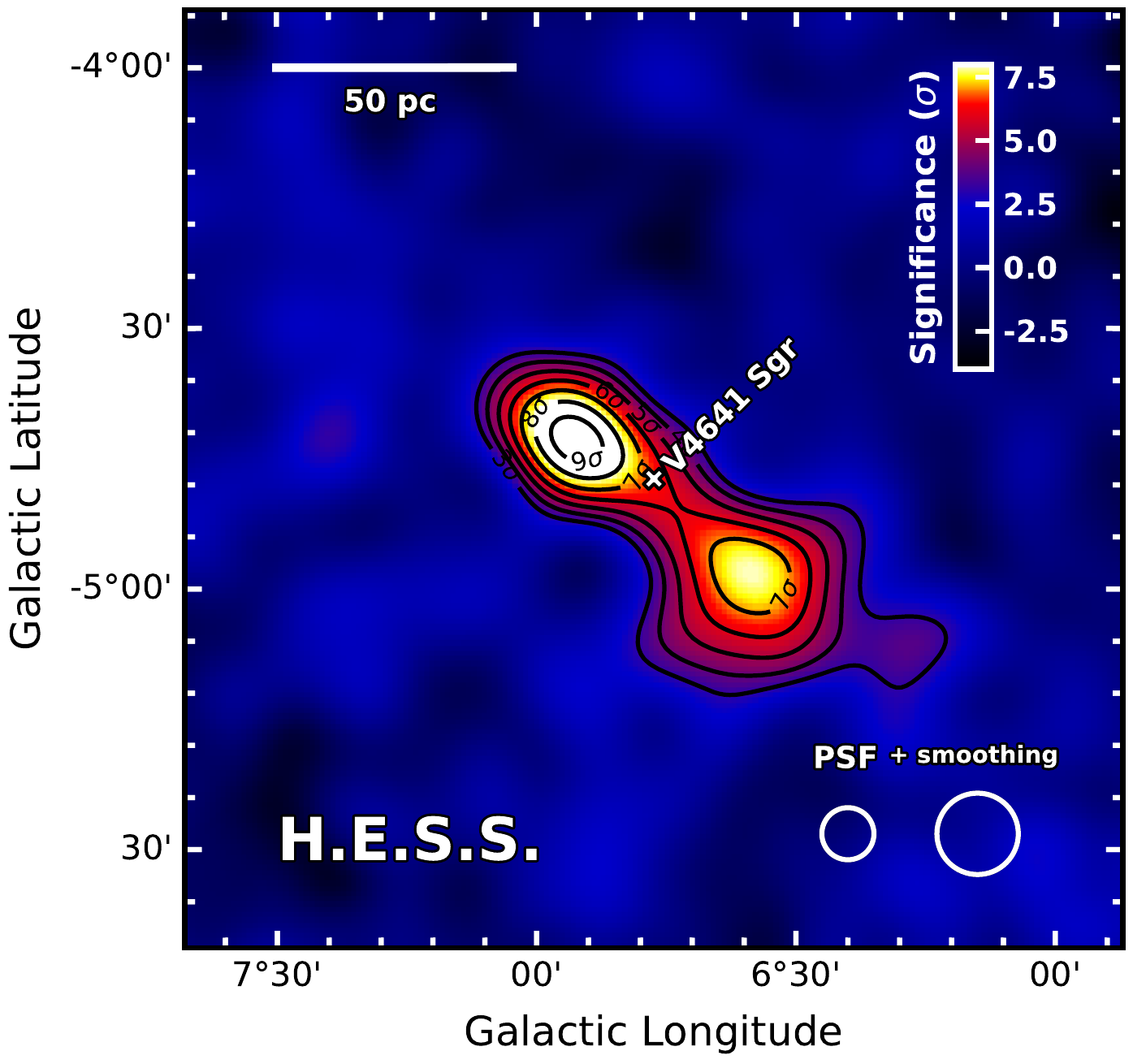}
	\caption{\label{fig:significance} Statistical significance of the \hess excess counts with energies higher than 0.8~TeV above the background of nearly isotropic cosmic rays (indicated by the colour scale) before statistical trials were accounted for. The 68\% containment region of the \hess PSF is denoted with a white circle (left). The map was derived using a maximum likelihood test for a Gaussian kernel of radius 0.06$\degree$, which resulted in smoothing. The scale resulting from the combination of the PSF and this kernel is shown with another white circle (right). }
\end{figure}

\begin{figure*}
	\centering
		\includegraphics[width=0.95\linewidth]{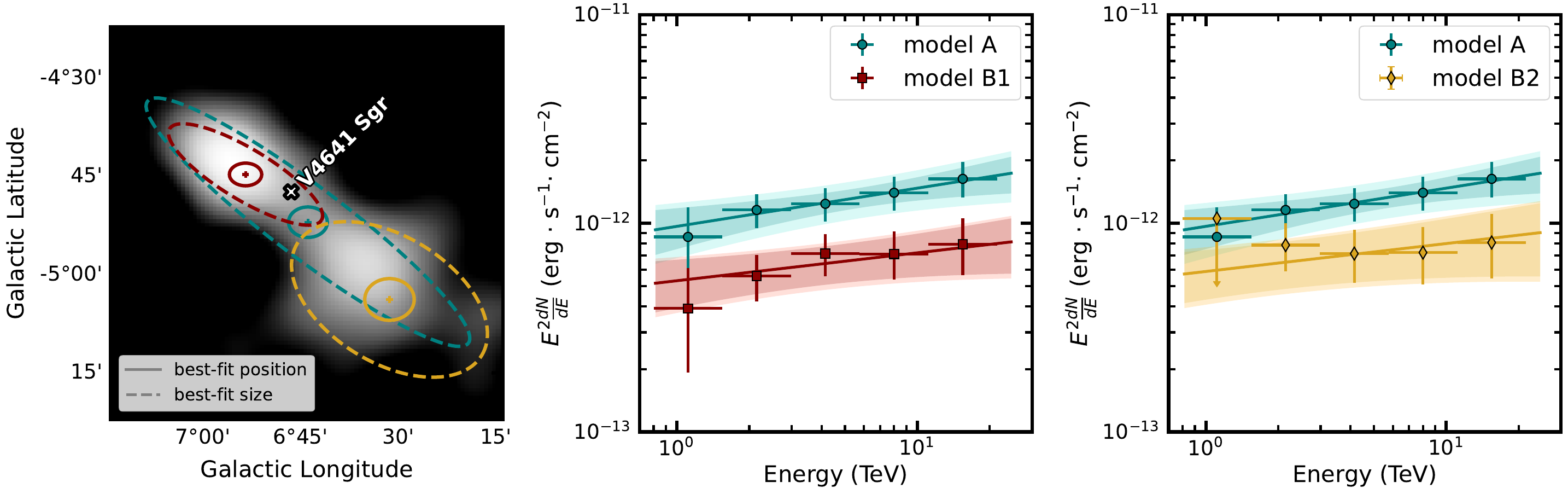} 
	\caption{\label{fig:spectra} \textit{Left:} Significance map. The 68\% containment regions (corresponding to 1.5$\sigma$ for a 2D Gaussian) of the spatial models are overlaid. The best-fit extent is depicted with dashed lines in teal for model A (see Section~\ref{subsec:morphology}), and in red and yellow for models B1 and B2 (see Section~\ref{subsec:morphology2}), respectively. The best-fit position and its 95\% confidence intervals are shown with crosses and solid lines using the same colour scheme.
    \textit{Middle:} The measured spectral energy distribution (SED) for models A (teal circles) and B1 (red squares). The solid line depicts the spectral shape of the best-fit power law. The dark and light shaded areas depict the statistical and systematic error bands, respectively. The error bars represent the combined systematic and statistical errors. The upper limits are shown at the 95\% confidence level.
    \textit{Right:} The symbols follow those of the middle panel, but for models A and B2 (yellow diamonds).}
\end{figure*}

\begin{table*}[b]
 \caption[]{\label{tab:fit_params} Best-fit parameters for the different spectro-morphological models.}
 \resizebox{\textwidth}{!}{\begin{tabular}{lccccccccc}
 \hline \hline
 \noalign{\vskip 1mm} 

  Model &  $\phi_0$ & $\Gamma$ & $E_0$ &    $l$ & $b$ & $\sigma_{\mathrm{major}}$  & $\sigma_{\mathrm{minor}}$ & $e$ & $\theta$\\
 &  ($10^{-13}$~TeV$^{-1}$cm$^{-2}$s$^{-1}$) &  &(TeV)  &  (deg) &  (deg)  &  (deg) &   (deg) &   & (deg)  \\ 

\hline
\noalign{\smallskip}
\multicolumn{9}{c}{ single elliptical component, TS$_A$=92320.849} \\
\noalign{\smallskip}
\noalign{\vskip 1mm} 

A   & $1.71^{\pm 0.27_{stat}}_{\pm 0.07_{syst}} $ 
& $1.82^{\pm 0.11_{stat}}_{\pm 0.05_{syst}}$ 
& $2$ 
& $6.731^{\pm 0.033_{stat}}_{\pm 0.005_{syst}}$
& $-4.866^{\pm 0.026_{stat}}_{\pm 0.005_{syst}}$ 
& $0.34^{\pm 0.04_{stat}}_{\pm 0.01_{syst}}$
&$0.06^{\pm 0.01_{stat}}_{\pm 0.01_{syst}}$
&$0.984^{\pm 0.005_{stat}}_{\pm 0.001_{syst}}$ 
& $53.0\pm 2.2$ \\
\noalign{\vskip 1mm}

\hline

\noalign{\smallskip}
\multicolumn{9}{c}{ two components, shared angle and index, ($\Delta$TS$_A$ = 19.8, $\Delta$N$_{dof}=5$ $\rightarrow$ 3.2$\sigma$)} \\

\noalign{\smallskip}
\noalign{\vskip 1mm} 

B1 & $0.91^{\pm 0.20_{stat}}_{\pm 0.10_{syst}} $ 
& $1.87^{\pm 0.11_{stat}}_{\pm 0.03_{syst}}$ 
& $2$ 
& $6.891^{\pm 0.027_{stat}}_{\pm 0.003_{syst}}$
& $-4.745^{\pm 0.019_{stat}}_{\pm 0.003_{syst}}$ 
& $0.15^{\pm 0.05_{stat}}_{\pm 0.01_{syst}}$
&$0.04\pm 0.02$
&$0.958\pm 0.037$ 
& $58.9^{\pm 4.6_{stat}}_{\pm 7.8_{syst}}$ \\
\noalign{\vskip 1mm} 

B2   & $1.01^{\pm 0.26_{stat}}_{\pm 0.08_{syst}} $ 
& - 
& $2$ 
& $6.523^{\pm 0.042_{stat}}_{\pm 0.005_{syst}}$
& $-5.063^{\pm 0.035_{stat}}_{\pm 0.005_{syst}}$ 
& $0.18^{\pm 0.05_{stat}}_{\pm 0.04_{syst}}$
&$0.11\pm 0.05$
&$0.810\pm 0.15$ 
& - \\
\noalign{\vskip 1mm} 

\hline

\end{tabular}}

\end{table*}

\section{Results}
\label{sec:the_results}

Figure~\ref{fig:significance} shows the significance map of the region around \vsgr in Galactic coordinates. An elongated excess is  detected, and the peak significance is 9.6$\sigma$. As a check of the background model, we constructed the background significance distribution by excluding the signal region with a circular mask with a radius of 0.7$\degree$ centred on \vsgr (Figure~\ref{fig:residual}). Fitting a Gaussian function to this distribution yielded a mean of $(-0.020\pm0.007)$ and a width of $(1.006\pm0.007)$. The deviation of the mean and width from 0 and 1 respectively translate to a marginal shift in the significance values: the corrected peak significance is 9.5$\sigma$.

\subsection{Morphology and spectrum of the emission}
\label{subsec:morphology}

The first and simplest model we considered (model A) is the combination of an elliptical Gaussian spatial model\footnote{\url{https://docs.gammapy.org/1.2/api/gammapy.modeling.models.GaussianSpatialModel.html}} with a power-law spectrum $\left(\frac{dN_{\gamma}}{dE}=\phi_0 \times \left(\frac{E}{E_0}\right)^{-\Gamma}\right)$. The choice to start with an extended elliptical morphology was motivated by the overall shape of the emission in the significance map. We fixed the reference energy $E_0$ to 2~TeV for this and all other models considered. This leaves seven free parameters: the spectral normalisation and index ($\phi_0$, $\Gamma$), the centre of the elliptical Gaussian ($l$, $b$), the semi-major axis size ($\sigma_{\mathrm{major}}$), the ellipticity ($e$), defined as $e = \sqrt{1 - \frac{\sigma^2_{\mathrm{minor}}}{\sigma^2_{\mathrm{major}}}}$, and the rotation angle of the major axis measured counter-clockwise from Galactic north ($\theta$). Including this source model in addition to the background model translates to a change in the test-statistic $\Delta$TS = 193.45 which, when accounting for N$_{\rm dof}=7$, corresponds to a significance of 12.9$\sigma$. The best-fit parameters of this model are summarised in Table~\ref{tab:fit_params}. The resulting spectrum and 68\% containment region  of the Gaussian are shown in Figure~\ref{fig:spectra}. The measured spectral index of $\Gamma=1.82\pm0.27_{\rm stat}\pm0.07_{\rm syst}$ is among the hardest measured for Galactic sources in this energy range~\citep{HESSCollaboration2018}. We found no evidence for curvature in the spectrum ($\Delta$TS=1.07) or an exponential cut-off ($\Delta$TS=1.01). The peak significance in the residual significance map produced after subtracting the model is $3.1\sigma$. The distribution of residual significance values are fitted by a Gaussian function of mean of $(-0.022\pm0.003)$ and a width of $(0.986\pm0.003)$, consistent with a well described signal and a well normalised background.

The single elliptical component has a major axis of $\sigma_{\mathrm{major}} = 0.34\pm0.04_{\rm stat}\degree$\footnote{In the following discussion of morphology parameters, we quote only statistical errors, see Table~\ref{tab:fit_params} for systematics errors}. The model is significantly ($5\sigma$) extended also in the minor axis direction: the best-fit ellipticity corresponds to $\sigma_{\mathrm{minor}} = 0.06\pm0.01_{\rm stat}\degree$. Accounting for distance, the 68\% containment radii of the spatial model (1.5$\sigma_{\mathrm{major/minor}}$ for a 2D Gaussian) correspond to  $\sigma_{\mathrm{major}} = (55.4\pm6.3_{\rm stat})d_{\rm 6.2kpc}$~pc and $\sigma_{\mathrm{minor}} = (9.8\pm2.0_{\rm stat})d_{\rm 6.2kpc}$~pc. We found that the position of the microquasar does not correspond with the best-fit centroid of the model: the angular distance between them is $0.124\pm0.032_{\rm stat}\degree$, which corresponds to a physical distance of more than 10$d_{\rm 6.2kpc}$~pc and which is inconsistent with zero by almost 4$\sigma$. In particular, the model centroid lies south (along the axis of the emission) of the position of the binary.

To test the assumption of a Gaussian morphology, we repeated the fit with an additional parameter $\eta$ which parametrises the shape between a flat elliptical disk morphology ($\eta=0.01$) and a peaked elliptical Laplace profile ($\eta=1$)\footnote{\url{https://docs.gammapy.org/1.2/api/gammapy.modeling.models.GeneralizedGaussianSpatialModel.html}}. A Gaussian shape is recovered when $\eta=0.5$. Freeing $\eta$ only marginally improved the fit, with $\eta=0.25\pm0.13_{\rm stat}$ preferred by 1.5$\sigma$. All other parameters fitted remained consistent within 1$\sigma$ uncertainties (see Table~\ref{tab:fit_params_appendix}). We concluded that the morphology is consistent with a shape in between an elliptical flat disk and an elliptical Gaussian, but we cannot distinguish between this scenario and the usually assumed Gaussian shape.

\subsubsection{Broad-band spectral energy distribution}
\begin{figure}
	\centering
		\includegraphics[width=0.97\linewidth]{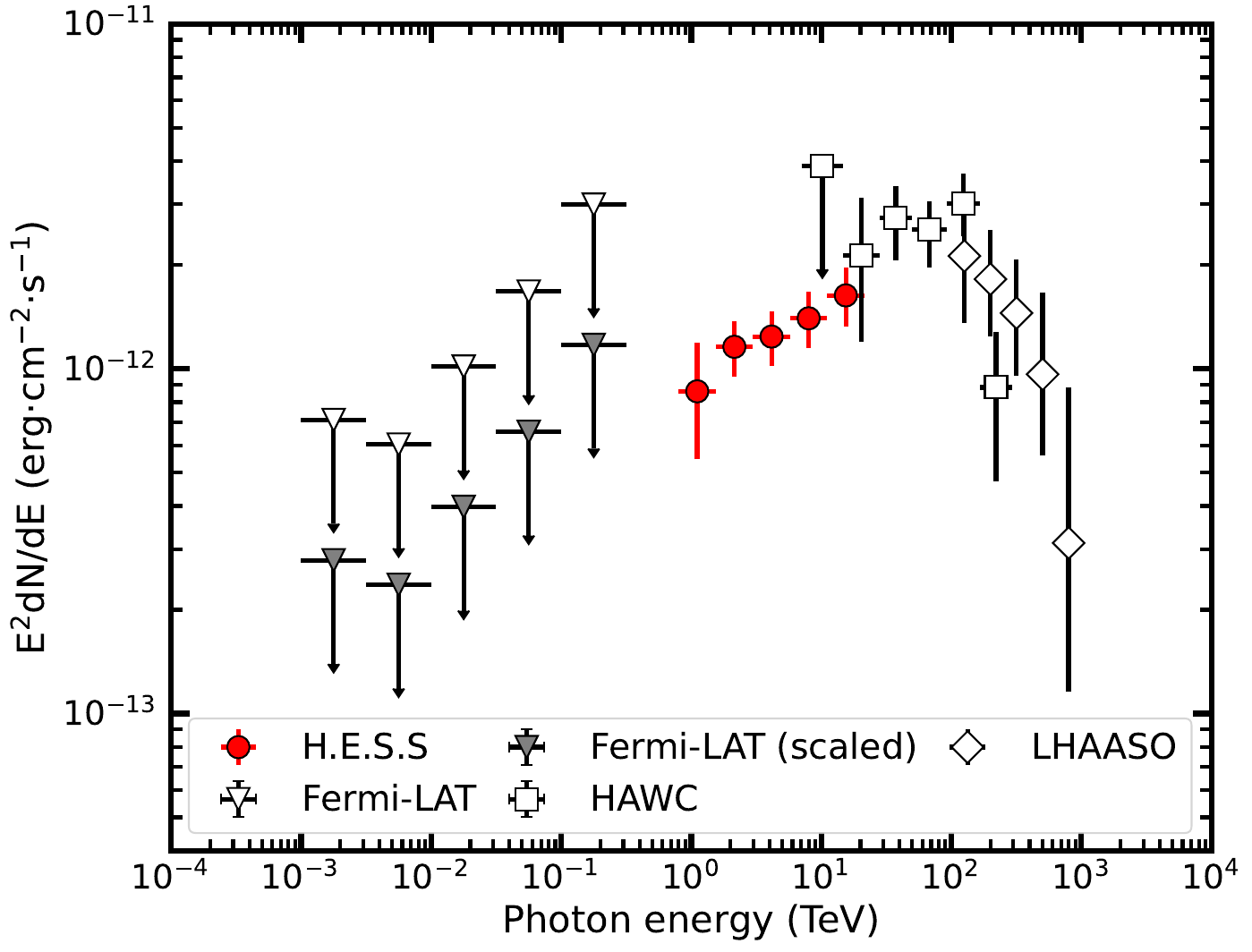} 
	\caption{\label{fig:sed} Broad-band spectral energy distribution. Flux points and upper limits measured by \hess (red circles), HAWC (squares), LHAASO (diamonds), and \fermi. The \fermi points are shown both as published in ~\cite{Zhao2025} (light triangles) and re-scaled to the \hess emission region size (dark triangles). The broad-band spectrum is observed to rise sharply until a peak forms at energies of $\approx$100~TeV, further extending up to energies of several hundreds of TeV.}
\end{figure}

The \hess spectral measurement and the existing measurements by \fermi~\citep{Zhao2025}, HAWC~\citep{hawcv4641} and LHAASO~\citep{lhaaso} reveal the full shape of the gamma-ray component in the spectral energy distribution. The \hess, HAWC and LHAASO spectral measurements connect smoothly, as can be seen in Figure~\ref{fig:sed}. The \hess data and the \fermi upper limits imply a rising spectral energy distribution in the 1-20~TeV range. HAWC and LHAASO data then show the turnover at energies around 100~TeV. All together, the data reveal, for the first time, a source spectrum which peaks at energies greater than 100~TeV.

In the GeV band we adopt the flux upper limits from~\cite{Zhao2025}, both as published and also rescaled down by a factor 2.6 to account for the fact that the extraction region in~\cite{Zhao2025} is not de-convolved with the LHAASO angular resolution, almost an order of magnitude larger than that of \hess The factor 2.6 is conservatively derived as the ratio between the angular size of the region with significance $>5\sigma$ in the LHAASO significance map and the 95\% containment region of model A (1326~arcmin$^2$).

\begin{figure*}[b]
	\centering
		\includegraphics[width=0.95\linewidth]{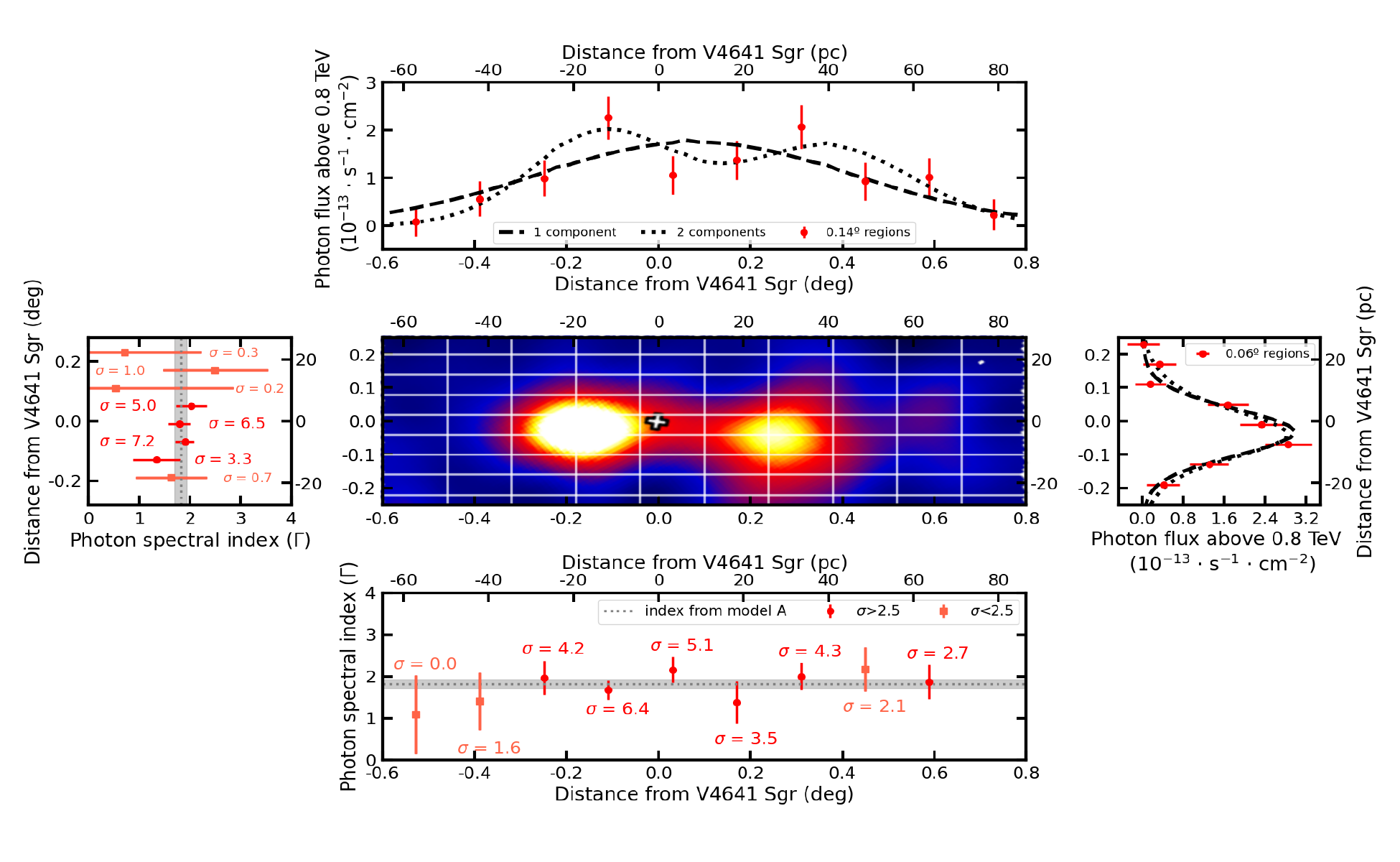} 
	\caption{\label{fig:profiles}Profiles along and across the emission. \textit{Centre:} Rotated and scaled version of the significance map is shown to help visualise the geometry of the different regions used to derive profiles. \textit{Top}: The red circles represent the measured flux above 0.8~TeV along the major axis as a function of distance to \vsgr. Distances are provided in degrees and parsecs adopting a distance of 6.2~kpc. The dashed and dotted lines depict the predictions of models A and B, respectively. \textit{Bottom}: The gamma-ray spectral index measured along the major axis in the same regions. Square and round symbols indicate when emission is detected inside the region with significance below or above $2.5\sigma$, respectively. A dotted grey line and shaded region indicate the best-fit value and statistical uncertainty of the index parameter in model A. The best-fit index at distance $\approx 0.7\degree$ is negative and is thus not visible in the plot. The significance of the emission in that region is 0.3$\sigma$. \textit{Middle left}: The symbols follow those of the bottom panel, but measured across the emission region.  \textit{Middle right}: The symbols follow those of the top panel, but measured across the emission region.}
\end{figure*}

\subsubsection{One or two components?}
\label{subsec:morphology2}

The significance map shown in Figure~\ref{fig:significance} suggests the presence of two peaks in the morphology of the emission, as also indicated by~\cite{hawcv4641}. We tested this hypothesis by adding a new component to the model, a second elliptical Gaussian spatial model with a power-law spectrum. We found weak statistical preference (3$\sigma$) for the inclusion of a second component. Among the different scenarios considered (see Tables~\ref{tab:fit_params} and~\ref{tab:fit_params_appendix}), the best statistical description when accounting for N$_{\rm dof}$ is two components with identical spectral index and orientation angle (model B, with components B1 and B2). Allowing more or all parameters to vary independently (models C and D, see Appendix~\ref{app:other_models}) did not improve the fit and resulted in statistically consistent best-fit parameters, in part because of increasing uncertainties. In particular, the ellipticity and inclination angle of the second component are poorly constrained when allowed to vary independently. We will consider the B model as the reference two-component scenario, but will take into account the spread of the parameters in models C and D when discussing physical implications. The best-fit parameters of models C and D are collected in Table~\ref{tab:fit_params_appendix}.

When considering two components, their 68\% containment radii in the major axes are $(24.5\pm8.2_{\rm stat})d_{\rm 6.2kpc}$ and $(29.7\pm7.5_{\rm stat})d_{\rm 6.2kpc}$~pc for the first and second component, respectively. Corresponding values for the minor axes are $(7.0\pm3.8_{\rm stat})d_{\rm 6.2kpc}$ and $(17.5\pm7.6_{\rm stat})d_{\rm 6.2kpc}$~pc. The extension along the minor axis is only weakly constrained in this scenario, with best-fit values of $\sigma_{\mathrm{minor}}$ consistent with zero within $\approx2\sigma$. In any case, a scenario with two point sources, as considered in~\cite{hawcv4641}, can be clearly excluded.

The relative sizes of the two components are marginally dependent on whether the orientation angle is allowed to differ between them, but are always consistent with each other within statistical uncertainties. We found that \vsgr does not lie halfway between both components, as one would naively expect in a symmetrical jet lobe scenario like that observed in \ssftt. Instead, it is much closer to the centroid of the B1 component than B2 ($0.125\pm0.019_{\rm stat}\degree$ vs $0.371\pm0.035_{\rm stat}\degree$), falling within the 68\% containment region of B1. This finding is consistent with the offset between the centroid of model A and the position of \vsgr discussed in the previous section. Similar distances were found for any of the other two component models considered (see Table~\ref{tab:fit_params_appendix}). The corresponding projected physical distances between \vsgr and the centroids in the models considered are in the range (10-15)$d_{\rm 6.2kpc}$ and (35-40)$d_{\rm 6.2kpc}$~pc for the first and second components, respectively. The centroids of the two components are separated by $0.486\pm0.040_{\rm stat}\degree$, which corresponds to over 50$d_{\rm 6.2kpc}$~pc.

In any scenario with two components, the overlap between them is non-negligible, leading to significant emission across the entire region, with no discernible gap between them. In particular, by fitting a modified version of model A with a gap centred on \vsgr of total height $h$, we were able to place an upper limit on how large this gap can be. We found that $h<0.06\degree$ (95\% containment), which translates to 6.5$d_{\rm 6.2kpc}$~pc. The same exercise around the centroid of model A yielded $h<0.15\degree$ (95\% containment), consistent with the weak preference for the two-component model.

 \begin{figure*}[b]
	\centering
		\includegraphics[width=.95\linewidth]{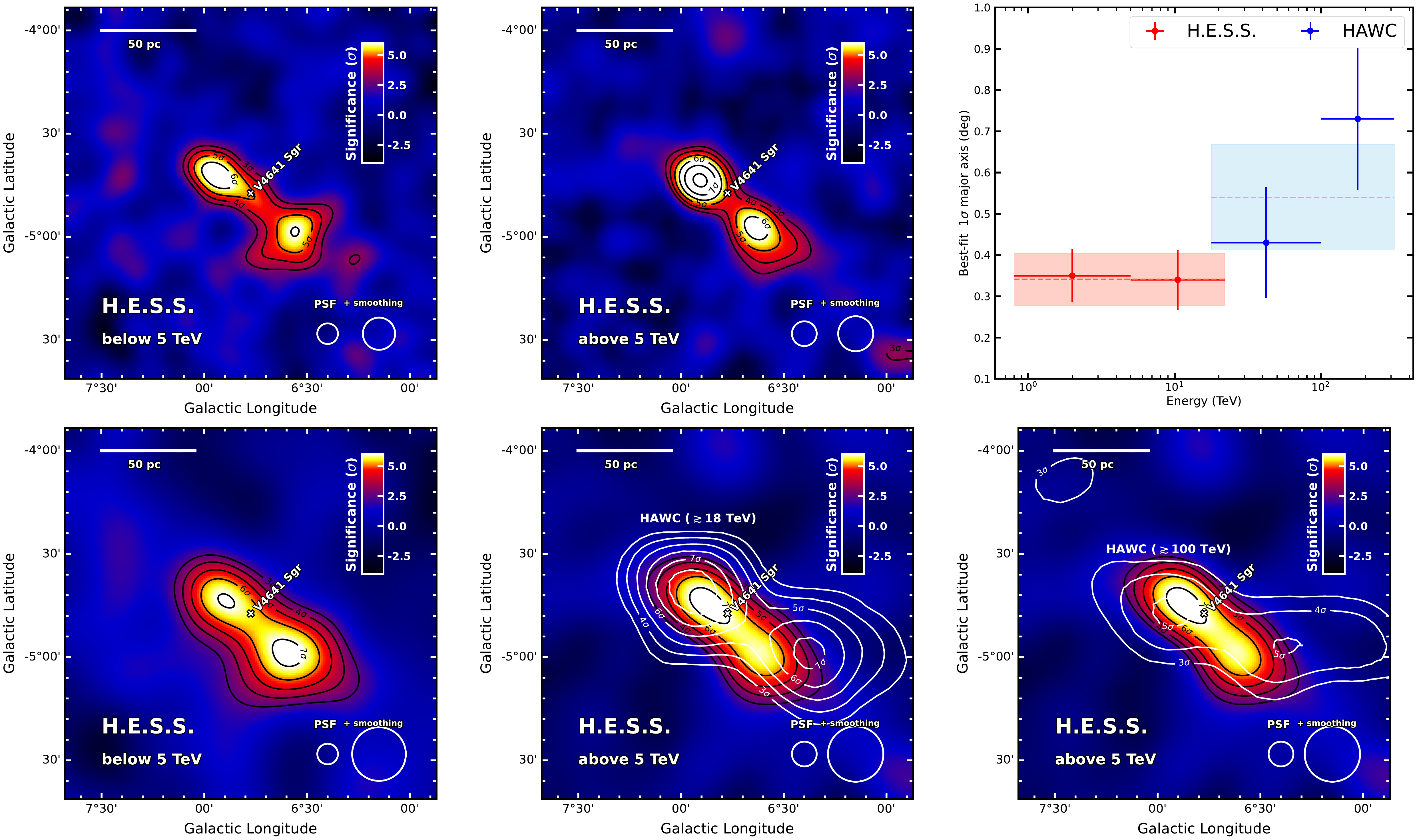}
	\caption{\label{fig:significance_energy} Significance maps in the energy bands. The symbols and colours follow those of Figure~\ref{fig:significance}, but for photon energies below 5~TeV (top left) and above (top middle). Significance maps derived with a larger smoothing kernel of 0.12$\degree$ are shown in the bottom row. The map for photon energies above 5~TeV is repeated to show the contours from the two HAWC significance maps provided in~\cite{hawcv4641}. Finally, the top right panel shows the size of the major axis in the "model A" fit for both \hess (red points) and for the HAWC publicly available data (blue points). The dashed lines and shaded bands represent the best-fit size and uncertainty derived in larger energy ranges as presented in Table~\ref{tab:fit_params} and~\cite{hawcv4641}. The error bars/bands represent the combined statistic and systematic uncertainty.}
\end{figure*}

\subsubsection{Spatial profiles}
To study the morphology in a model-independent way, we extracted the flux inside regions defined across and along the emission. The peculiar shape of the gamma-ray emission and the fact that \vsgr is not located at its centre require a clear definition of the along and across directions. The resulting profiles are one-dimensional and calculated as a function of distance to \vsgr. We defined both the across and along directions such that they pass through the microquasar position. We also derived profiles using the best-fit position of model A as a reference position and found the overall shape of the profiles to be highly consistent. We defined the direction we refer to as along the emission by fitting a combined spectral and spatial model like model A, but with the position fixed to that of \vsgr. This yielded $\theta = 55.06\pm3.33\degree$, which is consistent with the best-fit $\theta$ of model A (see Table~\ref{tab:fit_params}). The model with fixed position has $\Delta TS_A=-24.49$, which accounting for the extra two degrees of freedom rejects the model where the position is fixed to that of \vsgr by 4.6$\sigma$. The across direction is taken orthogonal to the along direction.

The fact that \vsgr is at the origin of the chosen coordinate system slightly obscures that in 2D, the position of \vsgr is not consistent with the centroid of model A or with the dip between models B1 and B2 as discussed above, due to projection effects. The central panel of Figure~\ref{fig:profiles} shows a rotated version of the significance map to aid the reader in visualising the geometry of the regions discussed. To measure the flux in each of the regions we fixed the spectral index to $\Gamma=1.8$, consistent with the index of the overall region. Further, we repeated the exercise with free $\Gamma$ to measure the spectral index in the different regions.

\paragraph{Flux and spectral index profiles along the emission}
 To measure the flux profile we used 10 rectangular regions of height 0.5$\degree$ ($\approx5\times\sigma_{\mathrm{minor}}$ of model A) orthogonal to the line defined by Galactic coordinates (7.2658$\degree$,-4.4455$\degree$) and (6.1181$\degree$,-5.2472$\degree$). Dividing this line into 10 sections translates to a region width of 0.14$\degree$, equivalent to more than the $95\%$ containment radius of the \hess PSF. The flux profile can be seen in the top panel of Figure~\ref{fig:profiles}. The shapes derived from Models A and B are also shown. The resulting flux was scaled to the solid angle size of the 10 regions used for the measurements. We did this for both model A and B and produced flux predictions. As can be seen, the flux profiles confirm the weak preference for a model with two components.

To measure the spectral index we used slightly different regions, of same width, but reduced height of 0.25$\degree$ (95\% containment of the minor axis of model A) in order to improve the signal-to-noise ratio. The procedure is the same as before, but with an additional free parameter (the index) in the fit. The best-fit index values can be seen in the bottom panel of Figure~\ref{fig:profiles}. We identified relevant regions with a significance threshold of $2.5\sigma$, and found that in all cases, the spectral index is consistent with that of the full region. No trend or pattern in the index can be discerned.

\paragraph{Flux and spectral index profiles across the emission}  We repeated the same procedure, but now with boxes of 0.06$\degree$ height and 1.2$\degree$ width. The flux and index profiles can be found in the left and right panels in Figure~\ref{fig:profiles}, respectively. The profile confirms that the emission is significantly extended in this direction (the \hess angular resolution is $\approx0.05\degree$), and again no trend or pattern can be discerned in the spectral index distribution, which is consistent with the average value.

\subsection{Energy-dependent morphology}
We explored the possibility of an energy dependent morphology in two ways: we split the data into two energy ranges: from 0.8 to 5~TeV and from 5 to 22~TeV (low and high energy, respectively) and 
 we repeated the profile estimation described above and the fits described in Section~\ref{subsec:morphology}. We found no evidence of energy-dependent morphology in either of these tests. Figure~\ref{fig:significance_energy} shows the significance maps derived in each of the energy ranges.
 
 In particular, we considered energy-dependent morphology in both the single and two-component model scenario. When fitting one elliptical component (with fixed spectral index) in the two energy bands, every parameter was consistent well within statistical uncertainties, both between energy ranges and with the fit across all energies. When considering the two-component model we also froze the angle $\theta$ and photon spectral index to those of model B. We found a weak hint that the first component becomes more symmetrical at higher energies, which can also be seen in Figure~\ref{fig:significance_energy}. The fit results have large statistical errors, however, with $e_{B1}^{>5~\rm{TeV}}=0.01\pm0.51$ when both components were fitted together (which translates into an upper limit of $e_{B1}^{>5~\rm{TeV}}<0.82$, $95\%$ C.L.) and $e_{B1}^{>5~\rm{TeV}}=0.67\pm0.36$ when the second component was frozen. All other parameters such as position or morphology at low energies remained consistent with those of model B. We conclude that there is no significant evidence for energy-dependent morphology in either the one or two component scenario.

While we do not find evidence of energy-dependent morphology within the \hess energy range, there might be when comparing our measurements to others made at a higher energies. In particular, the best-fit value of the major axis of model A ($\sigma_{\mathrm{major}}=0.34\pm0.04\pm0.01\degree$) is smaller than the one measured by HAWC~\cite[$\sigma_{\mathrm{major}}=0.54\pm0.08\pm0.1\degree$,][]{hawcv4641}, although consistent within 1.5$\sigma$ statistical and systematic uncertainties. Figure~\ref{fig:significance_energy} shows \hess significance maps above and below 5~TeV using both the usual kernel of $0.06\degree$ and a larger kernel of $0.12\degree$ which results in a total smoothing on the order of the angular resolution of HAWC, to facilitate visual comparisons between the instruments. Comparing these maps with the HAWC contours visually suggests a trend with energy.

To explore this possibility further, we made use of the publicly available dataset and analysis scripts provided by the HAWC Collaboration\footnote{\hyperlink{https://data.hawc-observatory.org/datasets/V4641/index.php}{https://data.hawc-observatory.org/datasets/V4641/index.php}} to repeat the fit with a single elliptical component reported by~\cite{hawcv4641}, but restricting the energy ranges to above and below 100~TeV. The best-fit $\sigma_{\mathrm{major}}$ for both the \hess and HAWC energy-restricted fits is shown in Figure~\ref{fig:significance_energy}. As can be seen, the extension measured with the HAWC data below 100~TeV ($\sigma_{\mathrm{major}}=0.43\pm0.09\pm0.1\degree$) is in better agreement with the size measured by \hess at lower energies. Above 100~TeV, however, the size almost doubles to $\sigma_{\mathrm{major}}=0.74\pm0.14\pm0.1\degree$. This is likely caused by the 3-4~$\sigma$ "tail" towards lower longitudes which can be seen in the HAWC contours above 100~TeV in Figure~\ref{fig:significance_energy}. Here we have assumed that the HAWC systematic uncertainty does not depend on energy. Nevertheless, even when comparing this measurement with that of \hess in the lowest energy range, the difference between them is less than 2.5$\sigma$.  We conclude that there is no significant evidence of energy-dependent morphology, neither in the \hess data alone nor when comparing with HAWC results.

\subsection{Time variability}
We searched for time variability in the region surrounding \vsgr with a focus on small spatial scales. The \hess observations probe timescales ranging between several minutes and several years - shorter than the light-travel time of a few parsecs. Consequently, we would only expect variability to occur on smaller regions and not the entire gamma-ray emission. We considered four regions of 0.07$\degree$ radius ($>$68\% containment for a point source, $\approx7d_{\rm 6.2kpc}$~pc): one centred on \vsgr, and three on the best-fit positions of models A and B. We derived light-curves in run-wise, monthly and yearly timescales. No variability is found in any of the regions, nor from a larger region that contains the entire gamma-ray emission.

 \begin{figure*}
	\centering
		\includegraphics[width=.95\linewidth]{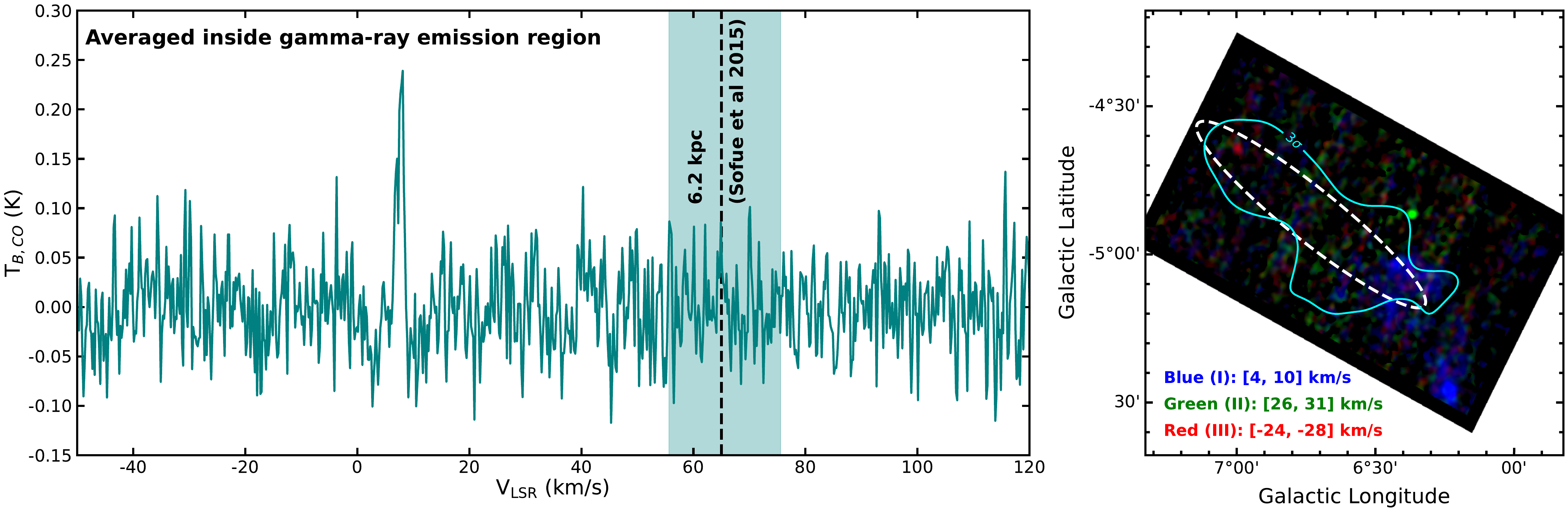}
	\caption{\label{fig:gas_main} Molecular gas around \vsgr. \textit{Left}: The brightness temperature profile of the CO emission averaged inside of the 68\% containment region of model A.  The shaded range indicates a range of 10~km~s$^{-1}$ around the velocity corresponding to the position of \vsgr. Note that this shaded band is for illustration purposes only, and that the density estimate reported in the text uses the noise level across the entire velocity range measured, regardless of what distance they correspond to. \textit{Right}: The RGB image depicting the integrated CO intensity in the three velocity ranges where small clouds are identified, unrelated to \vsgr. Smoothing with a Gaussian kernel of width 30~arcsec width was used to facilitate visualisation. The blue line shows the 3$\sigma$ contour of the \hess significance map (Figure~\ref{fig:significance}).The dashed white line shows the 68\% containment region of model A.  }
\end{figure*}

\subsection{Contributions from the central source}
Significant gamma-ray emission is detected from the position of \vsgr. When the extended components are not included in the fit, a point-like source model fixed to the microquasar coordinates provides a 4$\sigma$ improvement in the test-statistic; comparing the measured counts in a region of radius 0.1$\degree$ around \vsgr to the background reveals a significant (4.8$\sigma$) excess. When the contribution from the extended model components detailed in Table~\ref{tab:fit_params} is accounted for, however, the equivalent significance values drop to below 1$\sigma$. We derived upper limits for a possible additional contribution from the central source by including an additional point-source component fixed to the position of \vsgr in the model. The spectrum is assumed to be a power-law with a spectral index of 2. The fit is repeated, allowing the relevant parameters of the extended model(s) and the point-like source to vary. The result of this calculation is obviously impacted by the choice of model to describe the extended emission surrounding \vsgr. We derived upper limits using models A, B and D and find differences of at most 20\% between the resulting flux upper limits. Machine-readable upper limits (including likelihood profiles) are provided in the online material~\citep{HESSData} for model B, but for simplicity, and irrespective of the model, the 95\% differential flux upper limits ($E^2dN/dE$) between 0.8 and 22~TeV are contained within the range $(0.45-1.3)\times 10^{-13}$~erg~cm$^{-2}$~s$^{-1}$.

\section{Estimating the gas density around \vsgr}
\label{sec:the_gas_main}

A critical piece of information for the interpretation of the gamma-ray emission is the presence of dense gas in the surroundings which could act as a target for proton interactions. We used a combination of publicly available data products and dedicated observations with the Nobeyama Radio Observatory (NRO) to characterise the gas around \vsgr.

Dedicated $^{12}$CO(J=1-0) measurements were obtained with the NRO 45~m radio telescope between February and March 2025. An area of 0.8 $\times$ 1.4$\degree$ centred at R.A.=274.87$\degree$, Dec = 25.66$\degree$ was scanned in constant declination bands. The total observation time added up to roughly 20~h. The beamwidth of the NRO 45-m telescope is 15", a significant improvement compared to other existing CO observations of the region, such as the Dame 2022 survey~\citep{Dame2022}, with a beamwidth of 500". A few small clouds were identified, none at a distance consistent with \vsgr nor resembling the gamma-ray morphology. Figure~\ref{fig:gas_main} shows the velocity spectrum averaged inside the gamma-ray emission region, and an RGB map indicating the new clouds which were discovered. No molecular gas was found at a distance consistent with any estimates of the distance to \vsgr, regardless of the rotation curve used for the conversion.
Using the noise level in the data, we derived an  upper limit of n$_{\rm{gas}}<0.2$~cm$^{-3}$ for the average density in the gamma-ray emitting region (defined using model A). Details on the derivation of this value and validation using archival data are described in Appendix~\ref{app:gas}.

In addition to dedicated and archival CO observations, we considered the presence of atomic hydrogen through the study of 21-cm emission. We used the publicly available map from the HI4PI survey~\citep{Winkel2016} and followed the same procedure as for the CO maps, assuming optically thin gas. We found n$_{\rm{HI}}\approx 0.08$~cm$^{-3}$. The HI column density map shown in Figure~\ref{fig:gas} and the HI density estimate obtained differs significantly from those presented by~\cite{hawcv4641}. In that work, an integration velocity range three times larger than the average line widths of the warm neutral medium~\citep{Kalberla2018} was used. While the velocity range corresponds to a physical distance of almost 1~kpc~\citep[e.g.][]{Sofue2015}, all gas is assumed to be concentrated in a 20 pc depth, resulting in a higher density estimate ($\approx$1~cm$^{-3}$) than reported here. Further details and additional checks using H$\alpha$, free-free emission, and dust maps are presented in Appendix~\ref{app:gas}. Taken together, these observations reveal that \vsgr is located in a low-density environment. This finding is consistent with prior expectations given the fact that the scale height of Galactic molecular gas~\citep{Heyer2015} is several times smaller than the height of \vsgr below the Galactic plane. Finally, modelling of the XRISM data under a thermal assumption yielded n$_{\rm gas}< 0.3$~cm$^{-3}$~\citep{Suzuki2025}, further supporting our conclusion.

\section{Discussion}
\label{sec:the_discussion}

We have presented the \hess view of the gamma-ray emission around \vsgr. How this emission relates to the binary system and its jets is unclear. The reported low jet inclination angle would be expected to translate to a point-like source in analogy with the gamma-ray view of blazars, which differs significantly from the extended, asymmetrical emission we observed. In such a ``microblazar'' scenario in which particles are accelerated and emitting within a relativistic, low inclination jet, one would also expect one emission component to be enhanced by beaming, the other suppressed. The consistency between the flux of models B1 and B2 requires that, in such a scenario, emission arises from either a much slower jet termination region or interaction of escaped particles with their environment (i.e. large-scale field, molecular cloud).

\begin{figure}[h]
	\centering
		\includegraphics[width=0.45\linewidth]{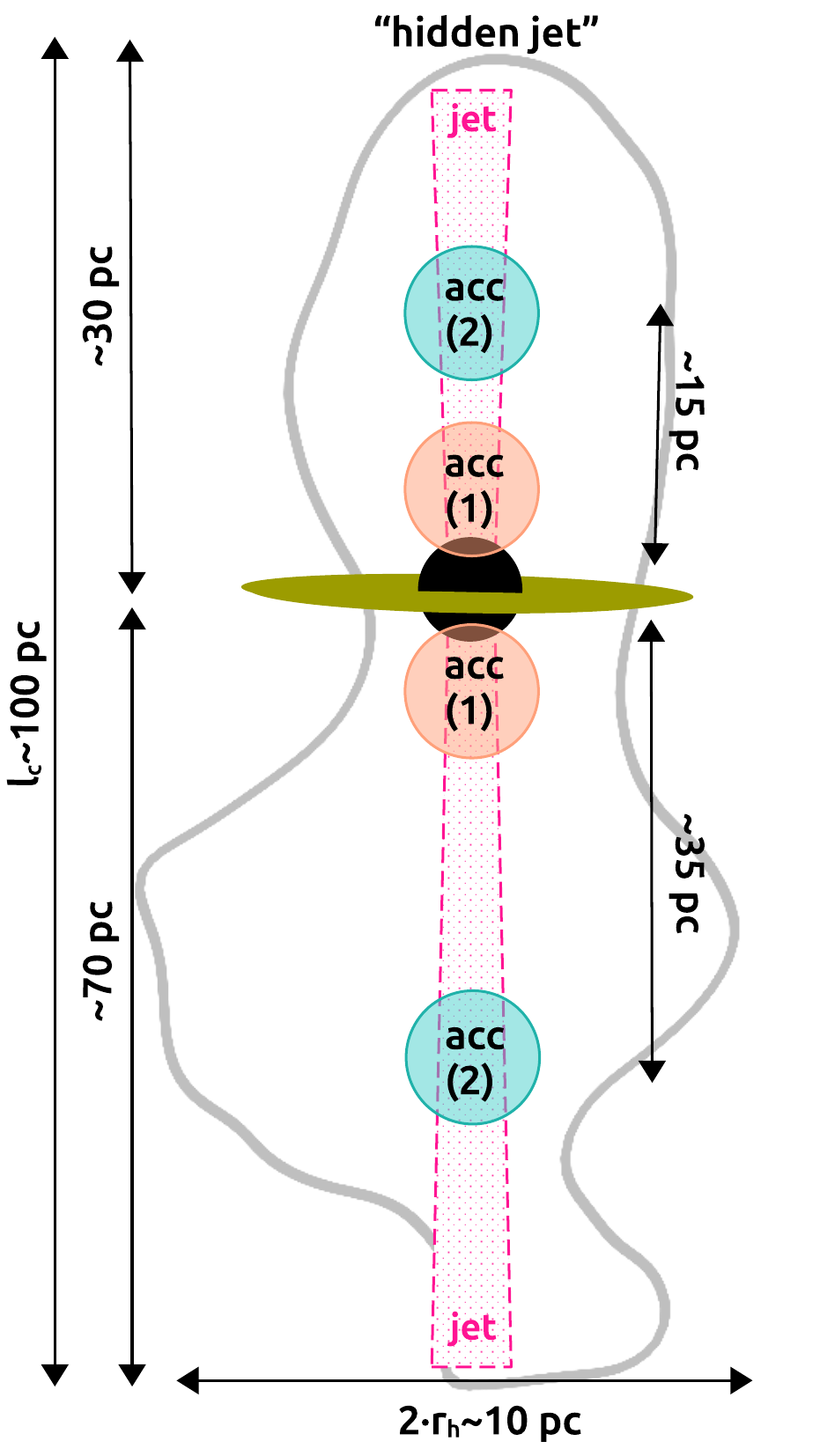}
		\includegraphics[width=0.45\linewidth]{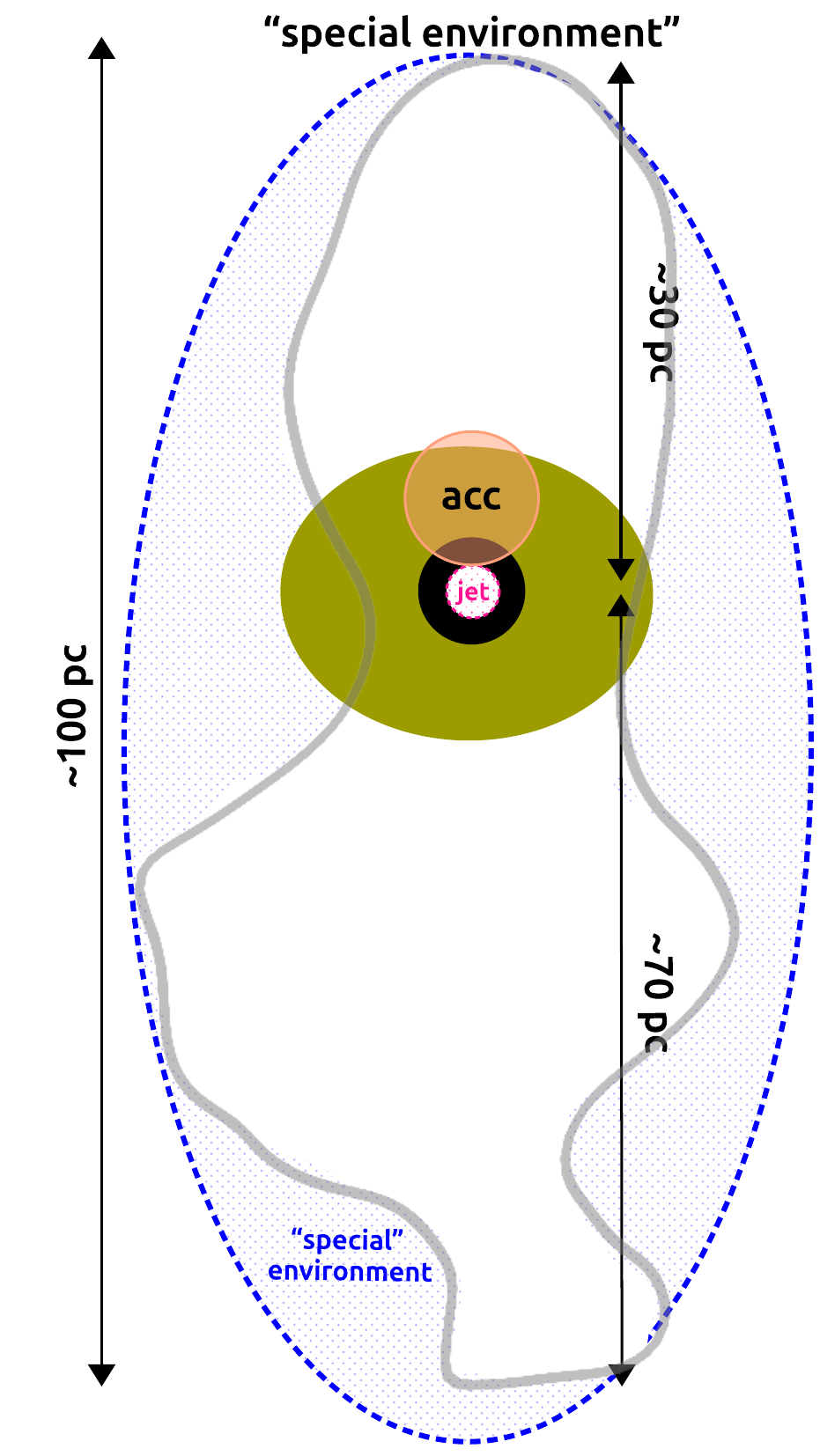}
	\caption{\label{fig:sketch} Cartoon depicting the two geometries. The gamma-ray emission region is outlined with a grey line that traces the rotated 3$\sigma$ contour of the map shown in Figure~\ref{fig:significance}. The black hole and accretion disc are shown assuming two options for the inclination angle: almost 90$\degree$ (left) or close to 0$\degree$ (right). In the perpendicular case, a pair of (hypothetical) jets launching from the black hole are depicted in magenta.  In the aligned jet case, a special environment that would explain the gamma-ray morphology is depicted in blue. The different acceleration sites discussed in the text are marked with circles. The sketch is annotated with scales derived from the best-fit gamma-ray models for the adopted distance of 6.2~kpc. }
\end{figure}

If the observed morphology is due to a nebula-like structure inflated by a jet~\cite[as discussed in][]{hawcv4641}, we can relate its size to the energetics of the central engine. We define an average jet power as $\bar L_{\rm jet}= \int_0^{t_{\rm jet}} L_{\rm jet} dt /t_{\rm jet} = \eta_{\rm jet} L_{\rm Edd}$, where $\eta_{\rm jet}$ is the jet efficiency.
Assuming the system did not move or reorientate during this period, one can follow the expressions in~\cite{Begelman1989} to estimate the time required to produce a jet-cocoon of full length $\ell_{\rm c}$ as

\begin{align}
\label{eq:tjet}
 t_{\rm jet}  \approx  \left(\frac{\ell_{\rm c}}{100\,{\rm pc}}\right) \left(\frac{r_{\rm h}}{5\,{\rm pc}}\right)
 \left(\frac{\eta_{\rm jet}}{0.1}\right)^{-{1\over 2}} \left(\frac{n_{\rm gas}}{0.2\,{\rm cm}^{-3}}\right)^{1\over 2}
 \left(\frac{v_{\rm jet}}{0.9c}\right)^{1 \over 2} {\rm Myr} ,
\end{align}

\noindent
where $r_{\rm h}$ is the jet cocoon head radius, which we take as 5~pc motivated by the measured $\sigma_{\mathrm{minor}}$ (see Fig. \ref{fig:sketch}). Note also that in Equation~\ref{eq:tjet} we adopt a jet velocity $v_{\rm jet}=0.9c$ following~\cite{Hjellming2000}, but that in principle, the velocity of the jet could be much lower in some of the scenarios considered in the following discussion.
We can compare this dynamical estimate with a characteristic time for mass-transfer in the binary system, taken as the thermal lifetime of the companion star $t_{\rm th}=GM^2/2RL$,  where $M$, $R$ and $L$ are its mass, radius and luminosity. Using numbers from \citet{MacDonald2014}, we find $t_{\rm th}\approx 100$ kyrs, though this number is known to provide underestimates~\citep{2000A&A...362.1046L, 2002ApJ...565.1107P}.

For the remaining discussion we will consider two competing scenarios with different implications for the acceleration and transport of particles in and around \vsgr.
\begin{enumerate}
    \item Jets are not aligned with the line of sight: A simple explanation for the elongated asymmetric emission is that there are jets with an orientation angle $i_{\rm jet}$ that is not aligned close to our line of sight. This scenario would be consistent with the reported orientation of the orbit and accretion disc~\citep{Gallo2014, Miller2002}, but inconsistent with the apparent super-luminal motion reported following the 1999 flare~\citep{Hjellming2000}. The left panel of Figure~\ref{fig:sketch} shows a cartoon summarising the proposed geometry. If the loss timescales of the gamma-ray emitting particles are sufficiently long, and confinement is effective, the emission could plausibly be a remnant of past activity, with the jet being currently inactive.  Observationally, this means that the gamma rays would be the best tracers of these hypothetical jets, since a low-density environment could hide evidence in other wavelengths. We will refer to this scenario as the ``hidden jet'' scenario in the following discussion. 

    \item Jets are aligned with the line of sight. Another possible explanation for the observed gamma-ray morphology is that the jets are indeed aligned with the line of sight, but that particles accelerated in the jets escape into a medium with properties that produce the elongated gamma-ray morphology. A simple example would be the presence of a molecular cloud with shape similar to the gamma-ray emission, which is illuminated by the particles accelerated by \vsgr. Alternatively the emission could signal the presence of a large-scale magnetic field configuration which traps the particles and leads to the observed morphology~\cite[as in, e.g. ][]{Neronov2024, Bao2024,Churazov2024}. The right panel of Figure~\ref{fig:sketch} shows a cartoon summarising the proposed geometry. We will refer to this scenario as the ``special environment'' scenario in the following discussion. Note that in this scenario, the extent and shape of the gamma-ray emission could be directly explained by the extent and shape of this special environment.
\end{enumerate}

In the hidden jet scenario, the jets launched by \vsgr have to extend around 50 pc to generate the observed morphology. In the aligned case, without resorting to a special environment and assuming e.g. $i_{\rm jet}$=10$\degree$~\citep{Hjellming2000} jets would have to extend more than 300~pc. As shown in Equation~\ref{eq:tjet}, producing 50~pc jets already poses tight constraints on the energetics, hence we do not elaborate further on this ``very extended jet'' case and require the special environment if the jet is aligned. Table~\ref{tab:werner} summarises all the considered scenarios and their feasibility.

\subsection{Possible acceleration sites}
While the existing observations are insufficient to pin-point the location of an acceleration site, they provide relevant constraints. We consider the two simplest scenarios motivated by the observations, with acceleration taking place either close ($\lesssim 3$~pc) to the binary system, in the corona or near the jet base, or further out in the jets, at distances of tens of pc. When transforming the angular size of the gamma-ray emission to physical distances throughout the discussion we will always assume (unless stated otherwise) an orientation orthogonal to the line of sight. Based on the uncertainties on the geometry of the system, however, every physical distance might be greater by a factor $(\sin \phi )^{-1}$, where $\phi$ is the unknown real inclination. We note that $\phi=i_{\rm jet}$ in the hidden jet scenario.

\begin{enumerate}
    \item Case 1: Acceleration close to the central source. If particles are accelerated somewhere near the compact object, in the corona or near the jet base~\cite[as in e.g.][]{Markoff2005, BoschRamon2005}, they would need to be transported out to $\approx$30~pc and $\approx$70~pc in each direction, respectively, while suffering minimal losses to avoid noticeable energy-dependent morphology. For simplicity, we will adopt $r_1=$50$\times (\sin{\phi})^{-1}$~pc as the minimum distance that particles need to travel in this scenario in the following discussion. This acceleration site location is applicable in both jet geometries considered (see Figure~\ref{fig:sketch}), although the dominant transport mechanism would likely be advection with the hidden jet flow or diffusion within the special environment, respectively.
    \item Case 2: Acceleration within the hidden jet. \ssftt is the only microquasar for which  the particle acceleration sites have been unequivocally identified~\citep{SafiHarb2022,Kayama2022,ss433}. In that case, the particle acceleration can be localised to positions along the jets, at a few tens of parsecs distance from the central binary. In analogy with \ssftt and motivated by the weak evidence for two components in the gamma-ray emission, we consider the possibility that the acceleration sites are located somewhere in the (hypothetical) hidden jet, or its lobes/cocoon. Assuming for simplicity that these acceleration sites are consistent with the centroids of models B1 and B2, this requires particles to be transported $\approx$15~pc and $\approx$35~pc in each direction - again for simplicity we will summarise this scale as $r_2=$25$\times (\sin{\phi})^{-1}$~pc. 
\end{enumerate}
The distinction between acceleration locations in cases 1 and 2 is most relevant in the hidden jet scenario: in the special environment scenario, particles would still have to travel a distance $\approx r_1$ to reproduce the morphology regardless of the site of acceleration within the jets (see Figure~\ref{fig:sketch}).

\subsection{Gamma-ray emission mechanism}
Translating the properties of the observed gamma-ray emission into constraints on particle acceleration and transport requires knowledge on which emission process is dominant. Electrons and positrons (hereafter referred to collectively as electrons) produce gamma-ray emission through IC scattering of soft photons to gamma-ray energies. The same electrons would also produce synchrotron emission at lower photon energies (radio to \xray), the intensity of which is  proportional to the energy density of the local magnetic field ($u_B$).  If instead hadrons (protons and heavier nuclei) are responsible for the bulk of the emission, they would produce gamma~rays via inelastic collisions with the surrounding gas. 

We will consider two competing scenarios, where the entirety of the gamma-ray emission is either produced by electrons or by protons. While a combined model is possible, we will not consider such a model here. There is no evidence for two components in the gamma-ray spectra and thus it would require fine-tuning while increasing the number of free parameters.

We used the open-source GAMERA package~\citep{Hahn2015, Hahn2022}, which includes Klein-Nishina corrections for the IC scattering calculations~\cite[e.g.][]{Blumenthal1970}, to model both the temporal evolution of the particle distributions and the resulting radiation. We compared the models to the observed broad-band spectra (Figure~\ref{fig:sed}). We fitted predictions from several one-zone emission models to this data. Since the likelihood profiles of the GeV upper limits from~\cite{Zhao2025} were not provided, they cannot be directly used in the likelihood fit and consistency with them was checked after fitting. For the model comparisons we included the \hess spectra with the flux points of Model A (Figures~\ref{fig:spectra} and \ref{fig:sed}) because (1) they correspond to the same spatial scale as the LHAASO measurements and (2) the spectral parameters of the two-component case are consistent, both for the analysis presented here and in the discussion in~\cite{hawcv4641}. Consequently, fitting the spectra separately does not add any additional information about the parent particle population. While the two-component model is suggestive of a scenario with two particle acceleration sites, the accreting binary remains the sole known source of power, meaning that any discussion about energetic requirements is unaffected. The existence of two acceleration sites is, however, considered when discussing constraints related to the morphology, as it effectively halves the distance that particles must travel to reproduce the observed extension. Figure~\ref{fig:sed} shows the prediction of the best-fit models in the leptonic and hadronic scenario, which are discussed in detail below.

\begin{figure*}
	\centering
		\includegraphics[width=0.9\linewidth]{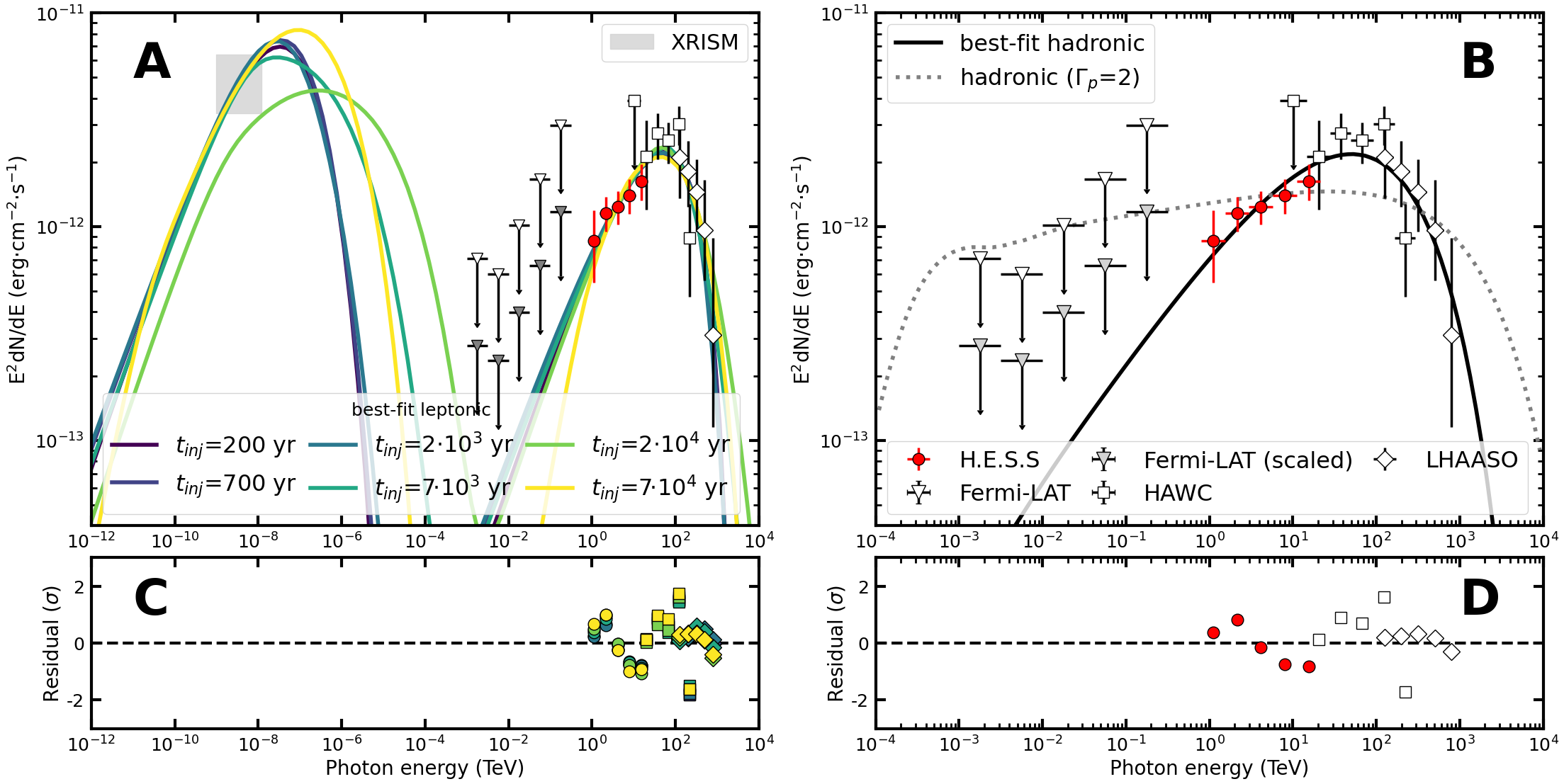}
	\caption{\label{fig:hadrolepto} Multi-wavelength SED and best-fit single-zones models. \textit{Panel A:} The flux points and upper limits follow those of Figure~\ref{fig:sed}. The XRISM flux region (scaled to the \hess emission region size) is depicted with a grey shaded region. The \fermi points are shown both as published in ~\cite{Zhao2025} (light triangles) and re-scaled to the \hess emission region size (dark triangles). The solid lines show the prediction of the best-fit leptonic model for different assumptions of $t_{\rm inj}$. \textit{Panel B:} The flux point symbols follow those of panel A. The black solid line shows the prediction of the best-fit hadronic model. The model obtained by fitting with a fixed $\Gamma_p=2$ is shown with a grey dotted line for reference. 
    \textit{Panels C and D:} The residuals between the data and the best-fit models. In the leptonic case, colours indicate the model and symbols the dataset.}
\end{figure*}

\subsubsection{Leptonic scenario}
\begin{figure}
	\centering
		\includegraphics[width=0.85\linewidth]{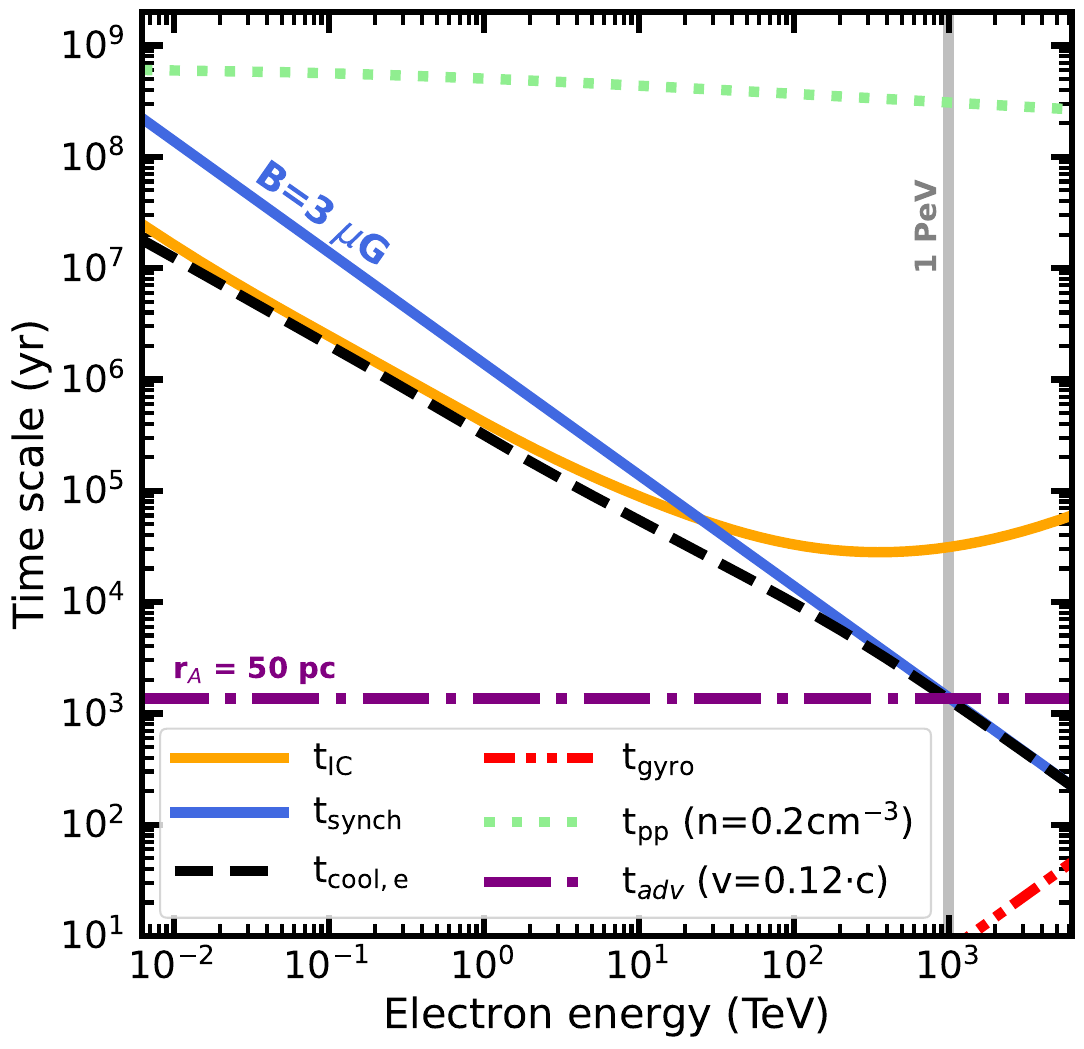}
	\caption{\label{fig:timescales} Relevant timescales. The cooling timescale as a function of electron energy (black dashed line) assuming $B=3$~$\mu$G and the same ambient radiation field as used in the broadband modelling. The contributions from IC and synchrotron are shown in orange and blue, respectively. The loss timescale of proton-proton interactions~\citep{Kafexhiu2014} is depicted with a dotted green line for density values of n$=0.2$~cm$^{-3}$. A solid vertical line indicates 1~PeV particle energy. The purple dashed-dotted line indicates the propagation time of a flow moving with $v=0.12c$, over 50~pc. The red dashed-double-dotted line shows the gyro-radius time defined as t$_{\mathrm{gyro}} = 2\pi r_g/c$ where $r_g$ is the particle gyro-radius for $B=3$~$\mu$G. For n$=0.2$~cm$^{-3}$ the bremsstrahlung loss timescale is $\approx 10^8$~yr, and thus only relevant for energies lower than those shown in this figure. }
\end{figure}

Relativistic electrons in low-density environments emit primarily via synchrotron and IC scattering. For relativistic electrons with Lorentz factor $\gamma$, the synchrotron loss time in a magnetic field with energy density $u_{\mathrm{B}} = B^2/8\pi$ is t$_{\mathrm{syn}}\propto (\gamma u_{\mathrm{B}})^{-1}$. 
For IC losses, the timescale is t$_{\mathrm{IC}}\propto (\gamma u_{\mathrm{rad}})^{-1}$, provided the Thomson regime applies, i.e. the typical target photon energy $\varepsilon_{\mathrm{ph}}$ satisfies $\gamma \varepsilon_{\mathrm{ph}}\ll m c^2$. Otherwise, Klein-Nishina suppression reduces the IC scattering rate, increasing the cooling time (see Figure~\ref{fig:timescales}).

To characterise the ambient radiation fields around \vsgr, we used the combination of a publicly available axisymmetric Galactic model for the energy densities of the diffuse interstellar radiation field~\citep{Popescu2017} and the Cosmic Microwave Background (CMB). The total energy density of the target photons is $u_{\mathrm{rad}}=2.2$~eV~cm$^{-3}$. We ignored the radiation from the companion star in the binary system because its energy density is negligible across the size of the gamma-ray emitting region.

\paragraph{Spectra:} Since the \xray synchrotron flux depends on $u_{\mathrm{B}}$, we could use the measurement by XRISM~\citep{Suzuki2025} to place an upper limit on the strength of the magnetic field. The short XRISM exposure was not sufficient to disentangle the nature of the \xray emission (synchrotron or thermal emission) or to accurately measure its spectral index. \cite{Suzuki2025} provided two options for the \xray spectra of the non-thermal emission: we combined both with a box between 1-12~keV ranging between $(1.6-3)\times 10^{-12}$~erg~cm$^{-2}$~s$^{-1}$ (see Figure~\ref{fig:hadrolepto}). We used this measurement as an upper limit to account for the possible thermal origin. The XRISM spectral extraction region is 629.4~arcmin$^2$~\citep{Suzuki2025}. We therefore scaled the measured flux to the 95\% containment region of model A (1326~arcmin$^2$). This approach is a priori conservative, since XRISM did not detect significant emission across the entirety of the gamma-ray emitting region that it covered, but only from a smaller region (see Figure~\ref{fig:xray}), suggesting that this upper limit could be lower. While there could in principle be bright \xray emission outside of the XRISM FoV, analysis of archival \textit{Chandra} and \textit{XMM-Newton} data with different coverage, partially covering regions outside the XRISM FoV, finds consistent flux measurements across the region~\citep{Suzuki2025}. The combined \textit{Chandra}, \textit{XMM-Newton} and XRISM observations cover 75\% of the gamma-ray emission region, defined as the spatial component of model A. Note that this estimate does not take into account differences in the acceptance across the FoV of the \xray instruments.

We modelled a population of electrons with spectra which follows a power-law of index $\Gamma_e$ with an exponential cut-off at $E_{\rm cut}$ (i.e. $\frac{dN_e}{dE_e}\propto E_{e}^{-\Gamma_e}\times\exp{(-\frac{E_e}{E_{\rm cut}^{e}})}$). Electrons were continuously injected for $t_{\rm inj}$, a time during which they lose energy due to interaction with assumed time-constant magnetic and radiation fields. The electron spectrum was normalised such that the integrated electron luminosity (above $E_{\rm min}^{e}$ = 1~MeV) is some fraction $\alpha_e$ of the Eddington luminosity of the system (L$_{\mathrm{Edd}}\approx10^{39}$~erg~s$^{-1}$). The value of $t_{\rm inj}$ has a direct impact on the spectral shape of the gamma-ray emission: the longer $t_{\rm inj}$, the more photons are produced at low energies, as a consequence of the accumulation of low-energy electrons due to the energy-dependent loss timescale. This introduces a degeneracy between the spectral index and the assumed $t_{\rm inj}$: the hard spectral index detected by \hess can be explained either with a low $t_{\rm inj}$ or a hard particle index. We perform the fit for several values of $t_{\rm inj}$ to assess the impact of this choice.

Owing to the computational cost of fitting the GAMERA model, we used a nested sampling algorithm~\citep{Skilling2004} to explore the multi-dimensional likelihood function and find its minimum. This approach has the advantage that it also samples the posterior distributions of the parameters, providing a more accurate representation of errors and correlation between parameters. We used the sampling interface included in \textit{Gammapy~2.0}~\citep{Donath2023, GammapyZenodo}, which derives posterior probability distributions and Bayesian evidence values through the nested sampling Monte Carlo algorithm
MLFriends~\citep{Buchner2016,Buchner2019} using the UltraNest\footnote{\url{https://johannesbuchner.github.io/UltraNest/}} package~\citep{Buchner2019}.

The best-fit values for different $t_{\rm inj}$ are shown in Table~\ref{tab:physical_models}. The parameters $\alpha_e$, $\Gamma_e$ and $E_{\rm cut}^{e}$ are highly correlated and as a consequence their posterior distributions cannot be described as Gaussian. For this reason, we provide the 68 and 95\% containment intervals (C.I.) for each of the fitted parameters to more accurately represent their range of validity. 

Regardless of the assumption of $t_{\rm inj}$, we found that the average magnetic field strength $B$ is $< 5~\mu G$ in all cases, with the best-fit values ranging between 2-3$~\mu G$. These values of $B$ should be interpreted as upper limits for a leptonic model, as we are assuming that the XRISM emission is entirely non-thermal and extends across the entire gamma-ray emission region. The obtained values of $B$ are an order of magnitude lower than the estimate for the jets of \ssftt~\citep{ss433}, which is the only other TeV microquasar for which the magnetic field strength has been estimated via a combined \xray and gamma-ray fit. 

While we are not able to distinguish statistically between the different $t_{\rm inj}$ considered (the largest $\Delta$TS<2), we can make physical arguments to constrain this parameter. Relatively high values of $t_{\rm inj}$ require either an extremely hard ($\Gamma_e<1$) particle spectral index or high ($E_{\rm cut}^{e}$>10~PeV) cut-off energy to explain the observed gamma-ray spectrum. For  $t_{\rm inj}\gtrsim10^4$~yr the likelihood surface begins to display two minima, each corresponding to one of these two scenarios. We chose to include in Table~\ref{tab:physical_models} an example of each with the $t_{\rm inj}=2\times 10^4$ and $t_{\rm inj}=7\times 10^4$~yr cases. While in principle such a hard spectral index or high-energy cut-off are not excluded, a scenario with $\Gamma_e\approx2$ and $E_{\rm cut}^{e}\approx$few~PeV is closer aligned with theoretical expectations for particle acceleration, and thus we interpret our results to tentatively suggest $t_{\rm inj} \lesssim 10$~kyr. Additionally, the preferred equipartition ratio of magnetic to electron energy density (which should be interpreted as an upper limit) is less than one for high values of $t_{\rm inj}$ (see Table~\ref{tab:physical_models}), which further disfavours such a scenario. Conversely, scenarios with a low $t_{\rm inj}$ require $>1\%$ of the Eddington luminosity to be injected into the electron population on average, a requirement which relaxes as $t_{\rm inj}$ increases. The resulting equipartition ratio in this case is greater than one for $t_{\rm inj}$ less than a few thousand years. Consequently, we suggest that $t_{\rm inj}$ is likely in the 1-10~kyr range, but note that this estimate is subject to several assumptions and thus uncertain. For $t_{\rm inj}$ in that range, we derive a lower limit on $E_{\rm cut}^{e}$ of $\approx 0.5$~PeV (95\% C.L.), with best-fit values around 1~PeV, with the caveat that this parameter is highly dependent on the particle index and thus the assumption of $t_{\rm inj}$.

\begin{table}
 \caption[]{\label{tab:physical_models} Best-fit values and confidence intervals (C.I.) for the parameters of the hadronic and a selection of the leptonic spectral models. }
\resizebox{\columnwidth}{!}{\begin{tabular}{lccc}
 \hline \hline
 \noalign{\vskip 1mm} 

  Parameter & Best-fit value &  68\% C.I. &  95\% C.I. \\
\noalign{\vskip 1mm} 

\hline
\noalign{\smallskip}
\multicolumn{4}{c}{ leptonic model, $t_{\rm inj}=2\times10^2$~yr} \\
\noalign{\smallskip}
\hline
\noalign{\vskip 1mm} 

$B$ ($\mu G$) & 2.72 & (2.11, 3.26) & (1.31, 3.82) \\
$\alpha_e$ (\% of L$_{\mathrm{Edd}}$) & 2.13  & (1.45, 5.13) & (1.10, 8.59) \\
$\Gamma_e$ & 1.85  & (1.70, 2.01) & (1.47, 2.07) \\
$E_{\rm cut}^{e}$ (PeV)   & 0.49 & (0.33, 0.86) &(0.21, 1.46)  \\
$u_{tot}^{e}$ (eV~cm$^{-3}$) & 0.08 & (0.06, 0.20)  & (0.04, 0.33) \\
$u_B/u_{tot}^{e}$  & 2.22 & - & - \\
\noalign{\vskip 1mm} 

\hline

\noalign{\smallskip}
\multicolumn{4}{c}{ leptonic model, $t_{\rm inj}=2\times10^3$~yr} \\

\noalign{\smallskip}
\hline
\noalign{\vskip 1mm} 

$B$ ($\mu G$)   & 2.87 & (2.26, 3.45) &  (1.48, 3.96) \\
$\alpha_e$ (\% of L$_{\mathrm{Edd}}$)   & 0.33 & (0.20, 1.41) & (0.13, 3.52) \\
$\Gamma_e$   & 1.93 & (1.82, 2.12) & (1.59, 2.19) \\
$E_{\rm cut}^{e}$ (PeV)   &  0.99 &  (0.59, 11.59) & (0.31, 68.41) \\
$u_{tot}^{e}$ (eV~cm$^{-3}$) & 0.13 & (0.08, 0.55)  & (0.05, 1.37) \\
$u_B/u_{tot}^{e}$  & 1.59 & - & - \\
\noalign{\vskip 1mm} 

\hline
\noalign{\smallskip}
\multicolumn{4}{c}{ leptonic model, $t_{\rm inj}=2\times10^4$~yr} \\

\noalign{\smallskip}
\hline
\noalign{\vskip 1mm} 

$B$ ($\mu G$)   & 1.91 & (1.54, 2.66) & (1.1, 3.5) \\
$\alpha_e$ (\% of L$_{\mathrm{Edd}}$)   & 0.04 & (0.03, 0.06) & (0.02, 0.1) \\
$\Gamma_e$   & 1.81 & (1.59, 1.87) &  (1.39, 1.98) \\
$E_{\rm cut}^{e}$ (PeV)   & 20.31 & (2.58, 49.09) &  (0.88, 90.95) \\
$u_{tot}^{e}$ (eV~cm$^{-3}$) & 0.16 & (0.12, 0.23)  & (0.08, 0.39) \\
$u_B/u_{tot}^{e}$  & 0.58 & - & - \\
\noalign{\vskip 1mm} 

\hline
\noalign{\smallskip}
\multicolumn{4}{c}{ leptonic model, $t_{\rm inj}=7\times10^4$~yr} \\

\noalign{\smallskip}
\hline
\noalign{\vskip 1mm} 

$B$ ($\mu G$)   & 2.84 &  (2.11, 3.42)& (1.4, 3.99) \\
$\alpha_e$ (\% of L$_{\mathrm{Edd}}$)   &0.03  &  (0.02, 0.09) &  (0.01, 0.19) \\
$\Gamma_e$   & $<$1.0& (1.0, 1.43) & (1.0, 1.62) \\
$E_{\rm cut}^{e}$ (PeV)   & 1.74 &  (1.45, 28.2)&  (1.16, 78.09) \\
$u_{tot}^{e}$ (eV~cm$^{-3}$) & 0.41 & (0.27, 1.22)  & (0.14, 2.58) \\
$u_B/u_{tot}^{e}$  & 0.49 & - & - \\
\noalign{\vskip 1mm} 

\hline
\noalign{\smallskip}
\multicolumn{4}{c}{ hadronic model} \\
\noalign{\smallskip}
\hline
\noalign{\vskip 1mm} 

$\Gamma_p$   &  1.49 &(1.24, 1.64)  & (1.04, 1.79) \\
$E_{tot}^{p}$($10^{50}\times$n$^{-1}_{0.2~cm^{-3}}$ erg)   &  8.0 & (7.17, 8.76) & (6.57, 10.31)  \\
$u_{tot}^{p}$ (n$^{-1}_{0.2~cm^{-3}}$ keV~cm$^{-3}$) & 0.49 & (0.44, 0.54)  & (0.40, 0.63) \\
$E_{\rm cut}^{p}$ (PeV) &  1.74 & (1.00, 2.69)  & (0.72, 4.99) \\

\noalign{\vskip 1mm} 
\hline
\end{tabular}}
\tablefoot{The normalisation parameter $\alpha_e$ is given as a percentage of L$_{\mathrm{Edd}}$ and not fraction. The energy densities of protons ($u_{tot}^{p}$) and electrons ($u_{tot}^{p}$) assume a cylindrical geometry, but we note that the depth of the emission region is completely unconstrained. Cooling and escape were ignored when deriving the energy densities. The magnetic field strengths should be interpreted as upper limits (see the discussion in the text) and, consequently, also the derived equipartition ratio $u_B/u_{tot}^{e}$. }
\end{table}

\paragraph{Morphology.} The energy-dependent loss timescale of electrons often equates to an energy-dependent morphology of the resulting emission \cite[e.g.][]{ Collaboration2019, ss433}. In particular, since the target radiation fields are isotropic and close to homogeneous on the spatial scales of interest, the morphology seen in IC gamma~rays reflects the combined action of cooling and particle transport, the latter also being energy-dependent. In the absence of a significantly energy-dependent morphology, two options are possible: either the combination of transport and energy loss is energy independent or energy-independent transport is faster than energy loss. Figure~\ref{fig:timescales} shows the relevant timescales as a function of energy. In order for a leptonic scenario to match the observed morphology, electrons must travel large enough distances without significant cooling.
As shown in Figure~\ref{fig:timescales}, synchrotron emission is the dominant loss mechanism for electron energies greater than 10~TeV, meaning that higher-energy particles lose energy faster than low-energy ones. In particular, for magnetic field strengths of 5 and 3~$\mu$G (motivated by the containment intervals reported in Table~\ref{tab:physical_models}) the energy loss times of a 1~PeV electron are respectively 500 and 1300~yr. 

If the jets are indeed aligned with the line of sight, transport out to large scales - in the special environment - would need to be diffusion-dominated, with the elongated morphology requiring the existence of a reasonably ordered large-scale magnetic field. The spatial extent of the emission can be governed by either cooling losses or by the extent of the "special environment" itself. 
In the first case, energy-dependent cooling and energy-dependent diffusion must conspire to avoid an energy-dependent morphology. In the second case, it is sufficient that cooling is slow enough to reach the edge of the "special environment" without significant losses. We can estimate how fast diffusion must operate in the cooling-limited case by taking $r_1 \approx \sqrt{2\times D(E) \times t_{cool}(E)}$, which for 1~PeV results in $D_{1~PeV} = (2.8- 7.6)\times 10^{29}$~cm$^2$~s$^{-1}$ for the range of magnetic field strengths considered. These diffusion coefficients are $\approx 25$ and 100 times larger than the Bohm diffusion coefficient at the same energy and value of $B$ and roughly one order of magnitude lower than the Galactic average value, consistent with estimates derived from observations of other gamma-ray sources~\citep[e.g. ][]{hawcgeminga}. 

The lack of observed energy dependence in the gamma-ray morphology between photon energies of 1 and 100~TeV also imposes constraints on the energy dependence of diffusive transport in the cooling-limited case. Adopting a diffusion coefficient $D(E)\propto E^{\delta}$, and cooling time $t_{cool}(E)\propto E^{-\gamma}$, we can write $r_1 (E)\propto E^{0.5(\delta - \gamma)}$. The ratio between $r_1$ for a difference of roughly two orders of magnitude in electron energy would then be $10^{\delta - \gamma}$. Conservatively requiring that $r_1$ does not change by more than a factor 2 in this energy range requires $|\delta - \gamma| < 0.3$. Figure~\ref{fig:timescales} shows that for electron energies in the relevant range ($\approx$4-400~TeV), the cooling time has a slope $\gamma \eqsim \frac{3}{4}$, which requires $ 0.45 \lesssim \delta \lesssim  1.05$. This constraint is dependent on the magnetic field strength, which could be even lower. For example if $B=1$~$\mu$G, IC would be the dominant cooling mechanism for electron energies slightly below 1~PeV, translating to a lower $\gamma$ which in turn constrains $\delta$ to a narrower range. Note that this estimate assumes that the emission region size is limited by cooling and not the extent of the special environment.

In the hidden jets scenario, a natural transport mechanism would be advection with the jet flow. Advection is energy-independent, so in order to avoid energy-dependent morphology along the jets like in the case of \ssftt~\citep{ss433}, advection must be faster than the energy loss timescale. This places a constraint $v>r_1/t_{cool}$ = 0.12$\times (\sin{\phi})^{-1}c$ for $B=3$~$\mu$G and $v>r_1/t_{cool}$ = 0.33$\times (\sin{\phi})^{-1}c$  for $B=5$~$\mu$G, where $\phi$ would be the jet inclination angle. Consequently, even assuming a relatively high magnetic field, this hypothetical advection speed would only need to be mildly relativistic to account for the observed gamma-ray morphology. The reported jet velocity of \vsgr is highly relativistic, but that is of course only relevant to the jet with a low inclination angle, which would not be consistent with the observed gamma-ray morphology. If the advection flow is indeed much faster than the constraints above and does not significantly decelerate, the emission region size would not be determined by cooling losses but instead it should be explained by the intrinsic size of the jets and/or their cocoon (see Equation~\ref{eq:tjet}). If we consider the situation where acceleration occurs at a point further along the jets (site 2; see Figure~\ref{fig:sketch}), the fact that there is significant gamma-ray emission in the region between the model centroids (see e.g.~Figure~\ref{fig:profiles}) rules out a pure advection model in that scenario. We cannot rule out that electrons might be accelerated at site 2, however, and instead of being advected further, they diffuse outwards from these locations. In that case the relevant spatial scale is $r_2$, and thus the diffusion coefficient would be a factor of 2 smaller than the values in the paragraphs above.

While low  magnetic field strengths ease energy losses in the leptonic scenario, they pose a challenge for accelerating particles to PeV energies. Assuming a single-zone configuration, in which the particle acceleration and emission occur under the same ambient conditions, following~\citet{Hillas1984}, we infer a maximum energy of

\begin{equation}
E_{\rm Hillas} \approx 3.5 Z \left(\frac{B}{3 \mu \mathrm{G}}\right) \left(\frac{v_{\rm jet}}{ c}\right)\left(\frac{R}{1\mathrm{pc}}\right)\mathrm{PeV},
\end{equation}
\noindent
where $R$ is the extent of the acceleration region. For $v_{\rm jet}=c$, that is, optimal acceleration conditions, an accelerator of $R\approx 1$~pc could produce 3.5~PeV electrons (in the absence of losses). For even lower magnetic field values (e.g. $B=1$~$\mu$G), this optimal accelerator would need to be larger than 1~pc to produce 1~PeV electrons. While these are extreme conditions, the acceleration sites in the jets of the microquasar \ssftt have been estimated to be $\approx 1-2$~pc~\citep{ss433, SafiHarb2022,Kayama2022}. There is a precedent for highly efficient acceleration in the Crab nebula~\citep{LHAASOCollaboration2021}.
Alternatively, there could also be a two-zone configuration, where particles are accelerated in a smaller region with a higher magnetic field, but manage to escape fast enough to a second zone where losses are less severe~\cite[e.g.][]{Breuhaus2021}.

\subsubsection{Hadronic scenario}
If protons are accelerated to relativistic energies somewhere in the \vsgr system, they can produce gamma-ray radiation through inelastic collisions with nearby gas ($pp$-interactions). As the energy dependence of the $pp$ cross-section far above threshold is approximately logarithmic, the energy loss timescale of protons has a weak energy dependence, as shown in Figure~\ref{fig:timescales}. The main parameter governing this process is the ambient density n$_{\rm gas}$, to which the energy loss rate is directly proportional. Conversely, the magnetic field strength is not well constrained in a hadronic scenario. Energy conservation limits the total energy emitted via synchrotron radiation from secondary electrons produced by decaying charged pions to not exceed that of the gamma~rays. Unfortunately, the XRISM \xray flux bounds are not sufficiently constraining, as they are higher than the measured gamma-ray flux.

\paragraph{Spectra.} We fitted the spectrum resulting from $pp$-interactions of a population of protons distributed following a power-law spectrum with an exponential cut-off: $\frac{dN_p}{dE_p}\propto E_{p}^{-\Gamma_p}\times \exp{(-\frac{E_p}{E_{\rm cut}^{p}})}$. As with the electrons, we normalised this power-law such that the integrated proton luminosity (above $E_{\rm min}^{p}$ = 1~GeV) is some fraction $\alpha_p$ of the Eddington luminosity of the system (L$_{\mathrm{Edd}}$). The value of $\alpha_p$ required to match the gamma-ray flux is inversely proportional to the value of n$_{\rm gas}$. We assumed that protons are injected continuously for $t_{\rm inj}$, where the only constraint on this time is that $t_{\rm inj}\lesssim 2$~Myr~\citep{Salvesen2020}. In order to avoid introducing further uncertainty, we discuss the proton normalisation in terms of the total energy, defined as $E_{\rm tot}^{p} = \alpha_p  \times L_{\mathrm{Edd}} \times t_{\rm inj}$ and simply note that the gamma-ray emission could require a higher or lower fraction of the jet luminosity depending on the assumed $t_{\rm inj}$.

We fitted the photon spectrum calculated with GAMERA using the same framework and gamma-ray data points as in the leptonic scenario. We found best-fit values for the index, cut-off energy and total energy of $\Gamma_p=1.49$, $E_{\rm cut}^{p}= 1.74$~PeV and $E_{\rm tot}^{p} = 8\times 10^{50} \times$ n$^{-1}_{0.2~cm^{-3}}$~erg, respectively. The 68 and 95\% containment intervals for each parameter are provided in Table~\ref{tab:physical_models}. The best-fit spectrum is shown in panel B of Figure~\ref{fig:hadrolepto} together with the measured flux points and upper limits. The cut-off energy is only weakly constrained towards higher energies, with $E_{\rm cut}^{p}\lesssim 5$~PeV (99\% containment limit). A particle index $\Gamma_p=2$ can be rejected with 5$\sigma$ certainty, as a hard particle spectrum is required by the spectral index measured with \hess and is consistent with the non-detection in the GeV band. It is generally thought that shock acceleration produces flat spectra with indeces in the 2-2.2 range~\citep{Drury1983} and while spectra as hard as 1.5 are theoretically possible~\citep{Kirk1989, Malkov1999, Summerlin2012}, they require extreme conditions or fine-tuning. Figure~\ref{fig:hadrolepto} also shows the best-fit spectra obtained when fixing $\Gamma_p$ to a value of 2. In this case, the required $E_{\rm cut}^{p}$ increases by almost an order of magnitude and the emission over-predicts significantly the GeV range while failing to describe the measured gamma-ray flux points.

The above energetic requirements pose a challenge for the hadronic scenario. As detailed in Section~\ref{sec:the_gas_main} and Appendix~\ref{app:gas}, there is no observed gas structure with a morphology resembling that of the gamma-ray emission (at any distance, or even combining all distances). The average density on the scale of the  gamma-ray emission is n$_{\rm gas}<0.2$~cm$^{-3}$, translating to $E_{tot}^{p} > 8 \times 10^{50}$~erg. If an average of $\approx$10\% of the jet power was injected as energetic protons, and the jet operated at L$_{\mathrm{Edd}}$ for $\approx$10\% of the time ($\alpha_p\times \eta_{\rm jet} = 0.01$), this process would need to be operating at this level for a minimum of 2.5~Myr. During this time, the source would have moved more than one degree in the sky~\citep{Salvesen2020}. Consequently, providing the observed gamma-ray flux requires extraordinarily efficient conditions.

\paragraph{Morphology.}  
The gamma-ray morphology in the hadronic scenario results from both the particles' spatial distribution and the shape of the target gas. As shown in Section~\ref{sec:the_gas_main}, there is no dense gas structure matching the observed gamma-ray morphology, which implies that its asymmetry would be the product of asymmetric transport. In the special environment scenario, having ruled out the presence of a matching molecular cloud, diffusion would need to be anisotropic and almost energy independent ($\delta \gtrsim 0$) in order to avoid resulting in observable energy-dependent morphology. 

Approximating the gamma-ray emission as a cylinder of radius 10~pc and height 110~pc results in a particle energy density of $u_{tot}^{p} > 0.5$~keV·cm$^{-3}$. Typical ISM conditions~\citep{Ferriere2001} can not confine such energy densities.
The protons might still be confined in a cocoon inflated by the jets, however. In this scenario, the morphology would be determined by the size of the cocoon, which follows Equation~\ref{eq:tjet}. This also requires that the proton confinement time, $t_{\rm conf}$ must be much greater than $t_{\rm jet}$, which in turn would be $t_{\rm jet}\gtrsim t_{\rm inj}$. Consequently, the jets would need to provide sufficiently strong magnetic fields to couple the particles to the cocoon material. Assuming equipartition with the protons gives $B=100\,\mu$G. A strict upper bound to the confinement time in the cocoon is then given by the Bohm limit ($D>D_{\rm Bohm}$, e.g.~\citet{Aharonian2004}):

\begin{align}
    t_{\rm conf} < \frac{R_c^2}{4 D_{\rm Bohm}} \approx 500\left(\frac{R_c}{50\,{\rm pc}}\right)^2 
    \left(\frac{E}{\rm PeV}\right)^{-1}
    \left(\frac{B}{100\,\mu{\rm G}}\right)\, {\rm kyr} 
\end{align}
 where $R_c \approx \ell_c/2$ is half the system size. While in principle the requirement $t_{\rm conf} \gg t_{\rm jet}$  can be fulfilled if, for example, the value of $v_{\rm jet}$ in Equation~\ref{eq:tjet} was lower or $\eta_{\rm jet}$ higher, a cocoon scenario would also imply even lower gas densities within the jet cavity. This in turn, exacerbates the extreme energy requirements described above, especially if the jet is only active for hundreds of kyr, which requires $\alpha_p\times \eta_{\rm jet} \gtrsim 0.1$.

We conclude that a hadronic origin of the gamma-ray emission is  unlikely. In such a low-density environment, not observing hadronic emission does not negate the existence of accelerated hadrons.

\section{Conclusions}
\label{sec:the_conclusions}
We have presented a detailed study of the gamma-ray emission around \vsgr, including physical modelling for a leptonic and a hadronic scenario. We reached a number of conclusions that we list below.
\begin{itemize}
\item The gamma-ray emission is highly elongated, with 68\% containment diameters of about 110$d_{\rm 6.2kpc}$ and 20$d_{\rm 6.2kpc}$~pc in either direction. The direction of this elongation disagrees with the proper motions of the source and with any other known intrinsic direction in the system.
\item The gamma-ray morphology shows weak evidence for two components. This is preferred by 3$\sigma$ over a model with a single elongated component. In any case, the spectra are consistent throughout the entire emission region, with a photon spectral index of $\Gamma\approx$1.8.

\item The spectral index measured by the \hess data implies a rising spectral energy distribution in the 1-20~TeV range. HAWC and LHAASO data then show a turnover at energies around 100~TeV. The data reveal a source spectrum that peaks at energies as high as $\approx$100~TeV for the first time. 

\item We found no evidence of an energy-dependent morphology in the 1-100~TeV energy range. Our comparison of the size of the emission across the major axis between \hess and HAWC at several energies~\citep{hawcv4641} showed consistency well within 1$\sigma$ across two orders of magnitude in energy. The size appears to increase above 100~TeV, but it remains consistent within the 2.5$\sigma$ uncertainties. Energy-dependent size measurements by the LHAASO Observatory at and above 100~TeV should be able to reveal whether this increase is real or simply a statistical fluctuation.

\item A leptonic scenario can explain the observed spectra if the magnetic field strength is $B<3~\mu$G. The timescale of the electron injection is not known and affects some of the best-fit parameters. We argued that $t_{\rm inj}$ is likely in the 1-10~kyr range by applying a number of assumptions. In this case, electrons above 1~MeV would require significantly less than 1\% of $L_{\mathrm{Edd}}$ to match the observed gamma-ray flux, with $E_{\rm cut}^{e}$ slightly below 1~PeV.
\item In the leptonic scenario, transport would need to be sufficiently fast for particles to cover dozens of pc without significant energy losses. We found that a hypothetical jet that is not aligned with the line of sight would only need to be mildly relativistic (v$_{\rm jet}\gtrsim$0.1c) to allow for this. If transport were instead dominated by diffusion, the elongated morphology would require a highly ordered weak magnetic field structure around the source. The lack of a strongly energy-dependent morphology constrains the energy dependence of the cooling-limited diffusive transport $D(E)\propto E^{\delta}$ to $ 0.45 \lesssim \delta \lesssim  1.05$.

\item Acceleration of $\approx 1$~PeV electrons in an environment with such a low magnetic field is  possible, but requires highly efficient acceleration and a relatively large (parsec-scale) acceleration site. These constraints would favour an acceleration site along the hypothetical jets, like in \ssftt~\citep{SafiHarb2022,Kayama2022,ss433}. Alternatively, electrons might be accelerated in a different region and be quickly transported to a region with a lower magnetic field.
\item Dedicated radio observations of the region and publicly available observations constrain the average ambient gas density to be low, with n$_{\rm gas}<0.2$~cm$^{-3}$. Lower values are found for neutral hydrogen.
\item A hadronic scenario with a cut-off energy of about 2~PeV can reproduce the observed spectra, but only with an unusually hard ($\Gamma_p$=1.5) proton spectral index and with a strong requirement for the proton power, given the low ambient density. In particular, we found that the gamma-ray flux requires the proton energy above 1~GeV to be $E_{tot}^{p} \gtrsim 8 \times 10^{50}$~erg, which corresponds to an energy density $u_{tot}^{p} \gtrsim 0.5$~keV·cm$^{-3}$. These numbers require at least 10\% of the average jet power of a jet, with a 10\% Eddington duty cycle being provided to protons for 2.5~Myr. This is the entire time for which the source is thought to have been at its current location~\citep{Salvesen2020}.

\item The high proton energy density requires extreme properties ($B>100~\mu$G) in order for the protons to remain confined in the gamma-ray-emitting volume long enough for the energy to accumulate. Given the distance of \vsgr to the Galactic plane and the lack of observational evidence for an unusual environment, these conditions require something other than the normal ISM. We considered a jet cocoon scenario, but found that in order to describe the observations, fine-tuning and extreme energetic requirements were imposed.

\item Because of these constraints, we conclude that the emission is most likely of leptonic origin. We propose the existence of a previously unknown large-scale outflow (hidden jet) aligned with the gamma-ray elongation axis, although a scenario with a highly ordered magnetic field structure (special environment) is also possible.

\item The observed elliptic morphology in the single and double component scenarios requires that regardless of the dominant emission process, particle transport must be inherently anisotropic, with a preferred direction along the major axis of the gamma-ray emission. At the same time, the observed asymmetry of the emission with respect to the position of \vsgr requires either an intrinsic warping or distortion of the jets and their cocoon or a specific magnetic topology that breaks the symmetry.

\item Our statement that the gamma-ray emission is most likely of leptonic origin does not imply that \vsgr is not a hadronic accelerator. Standard acceleration mechanisms do not distinguish between particles of different nature: If electrons are accelerated to energies above 1~PeV, protons might be as well.

\item In either scenario, the observed gamma-ray flux implies a luminosity of $L_{\gamma}\approx 5\times 10^{34}$~erg~s$^{-1}$, which is  hard to reconcile with the current luminosity of the system in other bands. Unlike \ssftt, which is known to have been in a super-Eddington accretion regime since its discovery, the only instance where the output of \vsgr was thought to have reached Eddington levels is the 1999 outburst. Since then, the source has been comparatively quiet, with only smaller outbursts with average luminosities of $\approx 10^{36}$~erg~s$^{-1}$. For example, during the 2020 outburst, the luminosity was estimated to be lower by about an order of magnitude than $L_{\mathrm{Edd}}$ even when obscuration was accounted for~\citep{Shaw2022}. The presence of bright extended gamma-ray emission on these spatial scales and the inferred maximum particle energies require that the system must have been consistently more active in the past \cite[e.g.][]{Peretti2025,2025ApJ...989L..25W}. While details depend on the scenario, the duration of this phase in the leptonic case with increased activity would need to be a few thousand years, while in the hadronic case, it would require timescales on the order of millions of years. This might exceed the thermal timescale of the companion star.

\end{itemize}
While the \hess observations (combined with those of \fermi, HAWC, and LHAASO) provide constraints on the physics that cause the emission around \vsgr, many questions remain. Upcoming \xray observations with \textit{Chandra}\footnote{\url{https://cxc.cfa.harvard.edu/target_lists/cycle27/cycle27_approved_peer_targets_by_radec.html}}, \textit{XMM-Newton}, and \textit{NuSTAR}\footnote{\url{https://heasarc.gsfc.nasa.gov/docs/nustar/ao11/c11_acceptarg.html}} will cover the entire gamma-ray emission region and will help us address some of them. For example, the \xray observations will potentially further validate the leptonic interpretation because detectable \xray emission is expected if $B>1~\mu$G. Additionally, in analogy with \ssftt~\citep{SafiHarb2022,Kayama2022}, the superior angular resolution of \xray instruments will be critical to constrain the location and nature of acceleration sites. Similarly, observations of the region with radio telescopes such as MeerKAT~\citep{MeerKAT} or the future Square Kilometre Array Observatory~\citep{SKAO} might provide a definite answer to the question of the jet geometry. In particular, recent radio campaigns have revealed the previously unknown large physical extent of jets originating in several \xray binaries~\cite[e.g. ][]{Wood2024, Motta2025}. Radio observations and a characterisation of the outflows from \vsgr would provide much needed clarity on the geometry of the system.

Observing \vsgr with future gamma-ray experiments such as the Cherenokv Telescope Array Observatory (CTAO)~\citep{CTAConsortium2019} or the Southern Wide-field Gamma-ray Observatory (SWGO)~\citep{SWGOCollaboration2025a} will also provide critical constraints. In particular, SWGO will be able to significantly improve the characterisation of the spectrum at the highest photon energies because \vsgr will transit overhead at the SWGO site, but only reaches the edge of the HAWC and LHAASO FoVs. Complementarily, the dramatic improvement in angular resolution that CTAO will provide at energies of dozens of TeV~\citep{Schwefer2024} will bridge the gap between the current \hess and HAWC/LHAASO morphological studies and either detect a weak energy-dependent morphology in the gamma-ray emission or rule it out entirely.

\vsgr was the second microquasar detected at energies above the TeV range, and similar to the first, \ssftt, a leptonic origin for the emission appears to be strongly preferred. The precessing jets of \ssftt are known to contain heavy nuclei~\citep[e.g.][]{Migliari2002}, however, and evidence for heavy nuclei in the jet of \vsgr has also been reported~\citep[e.g.][]{Zand2000}. There is no reason to expect that these particles are excluded from the acceleration process. While microquasars remain one of the most compelling candidates to account for multi-PeV Galactic cosmic rays, direct evidence of hadronic emission from these systems is still lacking. Given the sensitivity of current neutrino detectors~\citep{Neronov2024}, the most promising avenue remains the detection of gamma-ray emission coincident with dense gas near a microquasar. While the location of \vsgr in the Galaxy facilitates its association with the gamma-ray emission, the low-density environment would also hide evidence of hadronic acceleration. It is thus required to observe and detect emission from other microquasars in different environments, with the caveat that the source identification becomes difficult in crowded Galactic regions. One such promising example is the microquasar GRS~1915+105, which is located in a relatively dense region and from which GeV and TeV emission have recently been reported~\citep{MartiDevesa2025,lhaaso}. GRS~1915+105 is located in the Galactic plane nearby many putative gamma-ray sources, however, and a detailed study of the emission and its origin therefore likely requires observations with an improved angular resolution, such as that of the upcoming CTAO.

The gamma-ray emission around \vsgr is bright and persistent. It was detected by the HAWC Observatory as soon as improved algorithms expanded the HAWC FoV to include higher zenith angle ranges~\citep{Albert2024}. While \vsgr is only narrowly visible from the northern hemisphere, it transits almost overhead at the \hess site. The region was still observed by \hess for less than 10~h prior to the HAWC detection, which highlights the biases of pointed observations and the complementarity of IACTs and WCD arrays. Other sources like it may exist, but are simply located in poorly observed regions of the southern sky. This possibility would affect the expected contribution of microquasars to the cosmic-ray flux at petaelectronvolt energies. Additionally, the detection of emission surrounding \vsgr and other jetted \xray binaries~\citep{lhaaso} firmly establishes microquasars as a class of gamma-ray emitters. These findings require a re-evaluation of the nature of unidentified gamma-ray sources, some of which might be connected to a previously disregarded \xray binary.\\

\begin{spacing}{0.81}
{\raggedright \tiny \textit{Data availability:} The map shown in Figure~\ref{fig:significance} is available in electronic form at the CDS via anonymous ftp to \url{cdsarc.u-strasbg.fr} (130.79.128.5) or via \url{http://cdsweb.u-strasbg.fr/cgi-bin/qcat?J/A+A/}. Further data products, such as machine-readable versions of the flux points and models discussed in the paper are provided via Zenodo~\citep{HESSData}.}
\end{spacing}

\begin{acknowledgements}
We thank Hiromasa Suzuki for providing the XRISM image.    
The support of the Namibian authorities and of the University of
Namibia in facilitating the construction and operation of H.E.S.S.
is gratefully acknowledged, as is the support by the German
Ministry for Education and Research (BMBF), the Max Planck Society,
the Helmholtz Association, the French Ministry of
Higher Education, Research and Innovation, the Centre National de
la Recherche Scientifique (CNRS/IN2P3 and CNRS/INSU), the
Commissariat à l’énergie atomique et aux énergies alternatives
(CEA), the U.K. Science and Technology Facilities Council (STFC),
the Polish Ministry of Education and Science, agreement no.
2021/WK/06, the South African Department of Science and Innovation and
National Research Foundation, the University of Namibia, the National
Commission on Research, Science and Technology of Namibia (NCRST),
the Austrian Federal Ministry of Education, Science and Research
and the Austrian Science Fund (FWF), the Australian Research
Council (ARC), the Japan Society for the Promotion of Science, the
University of Amsterdam and the Science Committee of Armenia grant
21AG-1C085. We appreciate the excellent work of the technical
support staff in Berlin, Zeuthen, Heidelberg, Palaiseau, Paris,
Saclay, Tübingen and in Namibia in the construction and operation
of the equipment. This work benefited from services provided by the
H.E.S.S. Virtual Organisation, supported by the national resource
providers of the EGI Federation. We are grateful to the staff of the Nobeyama Radio Observatory (NRO) for their outstanding support during the 45-m telescope observations. The NRO is a branch of the National Astronomical Observatory of Japan (NAOJ), National Institutes of Natural Sciences.
\end{acknowledgements}

\bibliographystyle{aa}
\bibliography{V4641_Sgr}
\blfootnote{\label{current}{\large$^\star$}now at 29}

\begin{appendix} 

\section{Analysis details and derivation of systematics}
\label{app:extra_maps}
\subsection{Residual maps}
\begin{figure}[h]
	\centering
		\includegraphics[width=0.95\linewidth]{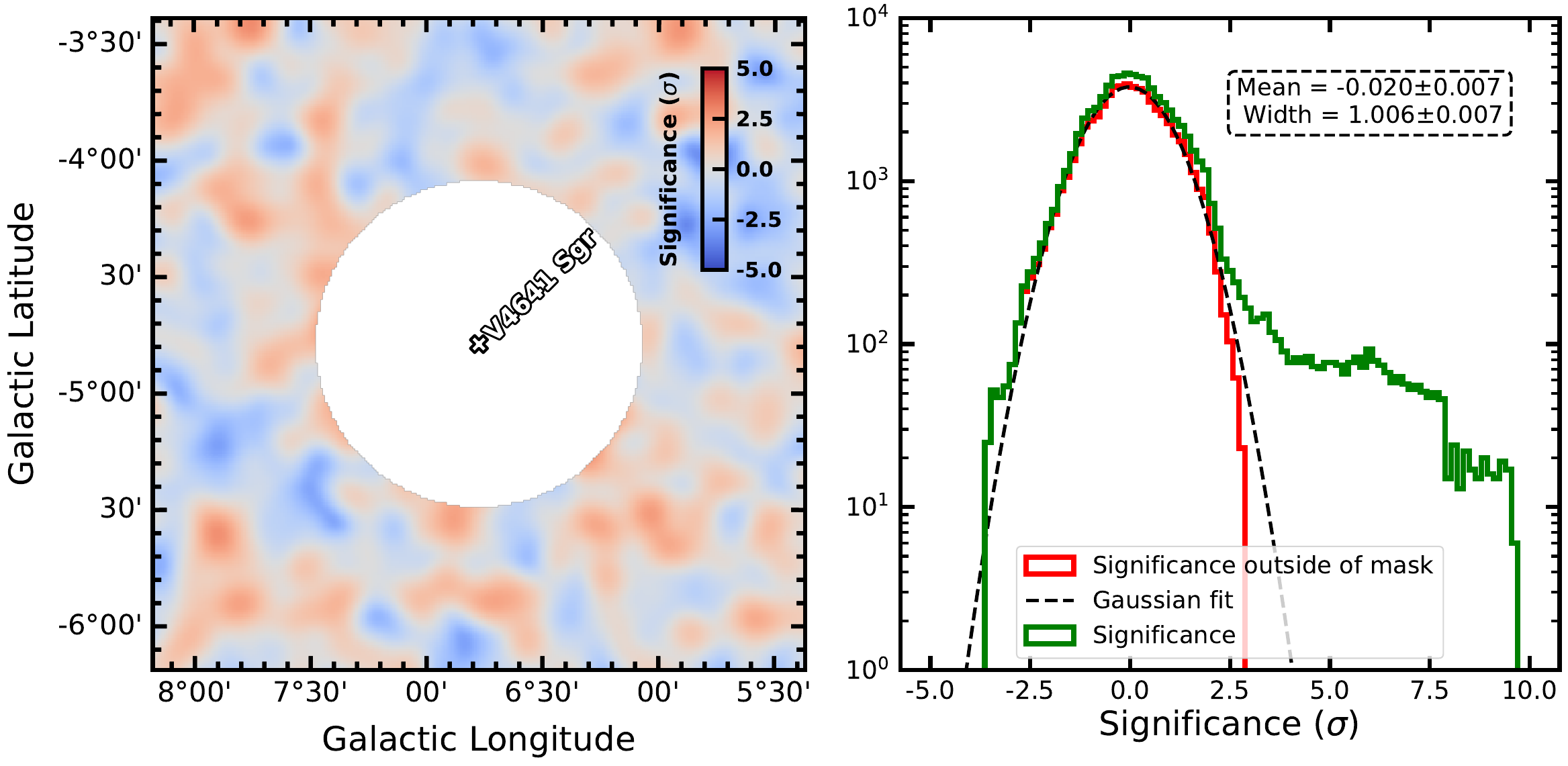}
	\caption{\label{fig:residual}  \textit{Left:} The same significance map as in Figure~\ref{fig:significance}, but with the exclusion mask covering \vsgr and a colour range which focuses on the background fluctuation scales. \textit{Right:} The distribution of the significance values outside of an exclusion mask covering \vsgr (red line) and over the entire map (green line). A Gaussian function is fit to the red distribution (black line), with mean $-0.020 \pm 0.007$ and width $1.006\pm 0.007$, close to the values of mean 0 and width 1 expected if the background is perfectly modelled.  }
\end{figure}

Figure~\ref{fig:residual} shows the same significance map as in Figure~\ref{fig:significance}, but with the emission region covered to highlight the background normalisation. The values in the map are used to make a histogram which is used to verify that the background model describes the observations well.

\vspace{-0.6cm}
\subsection{Additional models considered}
\label{app:other_models}
Table~\ref{tab:fit_params_appendix} contains the best-fit parameters of models which were considered, but which cannot be statistically distinguished from those in Table~\ref{tab:fit_params}. The model A$_\mathrm{disk}$ has an additional parameter $\eta$ which parametrises the shape of the spatial model between a flat disk morphology ($\eta=0.01$) and a peaked Laplace profile ($\eta=1$). The choice of shape has most impact on the extent of the emission for relatively high containment fractions. In the Gaussian scenario, the 68\% and 95\% containment radii along the major axis are $(55.4\pm 6.3)d_{\rm 6.2kpc}$ and $(88.6\pm 10.1)d_{\rm 6.2kpc}$~pc respectively, whereas allowing $\eta$ to be free results in $(59.5\pm 6.1)d_{\rm 6.2kpc}$ and $(70.2\pm 7.2)d_{\rm 6.2kpc}$~pc. 

Models C and D have two elliptical components with only one and no shared parameters, respectively. In both cases, the best-fit parameters are consistent with model B.

\begin{table*}
 \caption[]{\label{tab:fit_params_appendix} Best-fit parameters for additional spectro-morphological models considered.}
 \resizebox{\textwidth}{!}{\begin{tabular}{lccccccccc}

 \hline \hline
 \noalign{\vskip 1mm} 

  Model &  $\phi_0$ & $\Gamma$ & $E_0$ &    $l$ & $b$ & $\sigma_{\mathrm{major}}$  & $\sigma_{\mathrm{minor}}$ & $e$ & $\theta$\\
 &  ($10^{-13}$~TeV$^{-1}$cm$^{-2}$s$^{-1}$) &  &(TeV)  &  (deg) &  (deg)  &  (deg) &   (deg) &   & (deg)  \\ 

\hline
\noalign{\smallskip}
\multicolumn{9}{c}{ single elliptical component with $\eta=0.25\pm0.13$, ($\Delta$TS$_A$ = 2.2, $\Delta$N$_{dof}=1$ $\rightarrow$ 1.5$\sigma$)} \\
\noalign{\smallskip}
\noalign{\vskip 1mm} 

A$_\mathrm{disk}$   & $1.60\pm 0.24$ & $1.82\pm 0.11$ & $2$ & $6.718\pm 0.032$ & $4.880\pm 0.026$ & $0.56\pm 0.06$\tablefootmark{a} & $0.10\pm 0.02$\tablefootmark{a} & $0.983\pm 0.005$ & $52.6\pm 2.0$ \\
\noalign{\vskip 1mm}

\hline
\noalign{\smallskip}
\multicolumn{9}{c}{ two components, shared angle, ($\Delta$TS$_A$ = 20.1, $\Delta$N$_{dof}=6$ $\rightarrow$ 3$\sigma$)} \\
\noalign{\smallskip}
\noalign{\vskip 1mm} 

C1  & $0.88\pm 0.21$ & $1.80\pm 0.16$ & $2$ & $6.889\pm 0.028$ & $-4.746\pm 0.20$ & $0.16\pm 0.05$& $0.04\pm 0.02$ & $0.961\pm 0.033$ & $58.6\pm 4.4$ \\
\noalign{\vskip 1mm} 

C2   & $1.08\pm 0.29$ & $1.96\pm 0.19$ & $2$ & $6.522\pm 0.042$ & $-5.065\pm 0.036$ & $0.18\pm 0.05$  & $0.11\pm 0.05$& $0.792\pm 0.174$ & -\\
\noalign{\vskip 1mm} 

\hline
\noalign{\smallskip}
\multicolumn{9}{c}{ two components, all free, ($\Delta$TS$_A$ = 21.8, $\Delta$N$_{dof}=7$ $\rightarrow$ 3$\sigma$)} \\
\noalign{\smallskip}

\noalign{\vskip 1mm} 

D1  & $0.68\pm 0.19$ & $1.75\pm 0.16$ & 2 & $6.917\pm 0.023$ & $-4.726\pm 0.016$ & $0.11\pm 0.03$ & $0.05\pm 0.02$ & $0.894\pm 0.085$ & $60.2\pm 9.1$\\
\noalign{\vskip 1mm} 

D2   & $1.12\pm 0.25$ & $1.95\pm 0.18$ & 2 & $6.543\pm 0.03$ & $ -5.014\pm 0.026$ &  $0.17\pm 0.07$ &$0.10\pm 0.07$ & $0.815\pm 0.261$ & $59.3\pm 19.7$\\
\noalign{\vskip 1mm} 

\hline
\noalign{\vskip 1mm}

\end{tabular}}
\tablefoot{Uncertainties are statistical only. Systematic uncertainties should be taken as of similar magnitude as those in Table~\ref{tab:fit_params}. \tablefoottext{a}{Note that the best-fit spatial scales $\sigma_{\mathrm{major}}$ and $\sigma_{\mathrm{minor}}$ correspond to roughly 68\% containment radii for $\eta=0.25$ but 39\% for the Gaussian ($\eta=0.5$) case.}}
\end{table*}

\subsection{Derivation of systematics}
\label{app:systematics}
We estimate systematic uncertainties in the model parameters with a Monte Carlo-based approach following~\cite{Collaboration2023}. The concept behind this approach is that there might be discrepancies between the instrument response functions (IRFs) associated to each run and the real conditions of observations, which would introduce systematic biases in the analysis results. We identify two main sources of uncertainty: mis-modelling of the background and shifts in the energy reconstruction caused by changing atmospheric conditions. 

Using a Monte Carlo approach, we randomly modify the IRFs to account for the two effects mentioned above, and use these modified response functions to simulate observations following the best-fit source models. We then re-fit these pseudo-dataset using the original IRFs to mimic the scenario where the IRFs do not match the recorded data. The obtained spread in the fitted model parameters then reflects their combined statistical and and systematic uncertainty. For further details on the ways the IRFs are modified and how the scale of the modifications is chosen we refer to ~\cite{Collaboration2023}. For certain fitted parameters, such as the coordinates and position angle of the elongated Gaussian components, the distribution is not influenced by the systematic effects considered in this scheme. However, these parameters may be affected by other systematic effects neglected in the Monte Carlo  approach. In particular, the source positions are subject to a systematic uncertainty of the pointing position of the \hess telescopes, which is of the order of 10" - 20"~\citep{GillessenThesis} and negligible compared to the statistical errors.

The same pseudo-datasets are used when calculating systematic errors for flux points. In this case, the systematic uncertainties are negligible compared to the statistical ones.

\vspace{-0.2cm}
\section{\xray coverage}
\begin{figure}[h]
	\centering
		\includegraphics[width=0.9\linewidth]{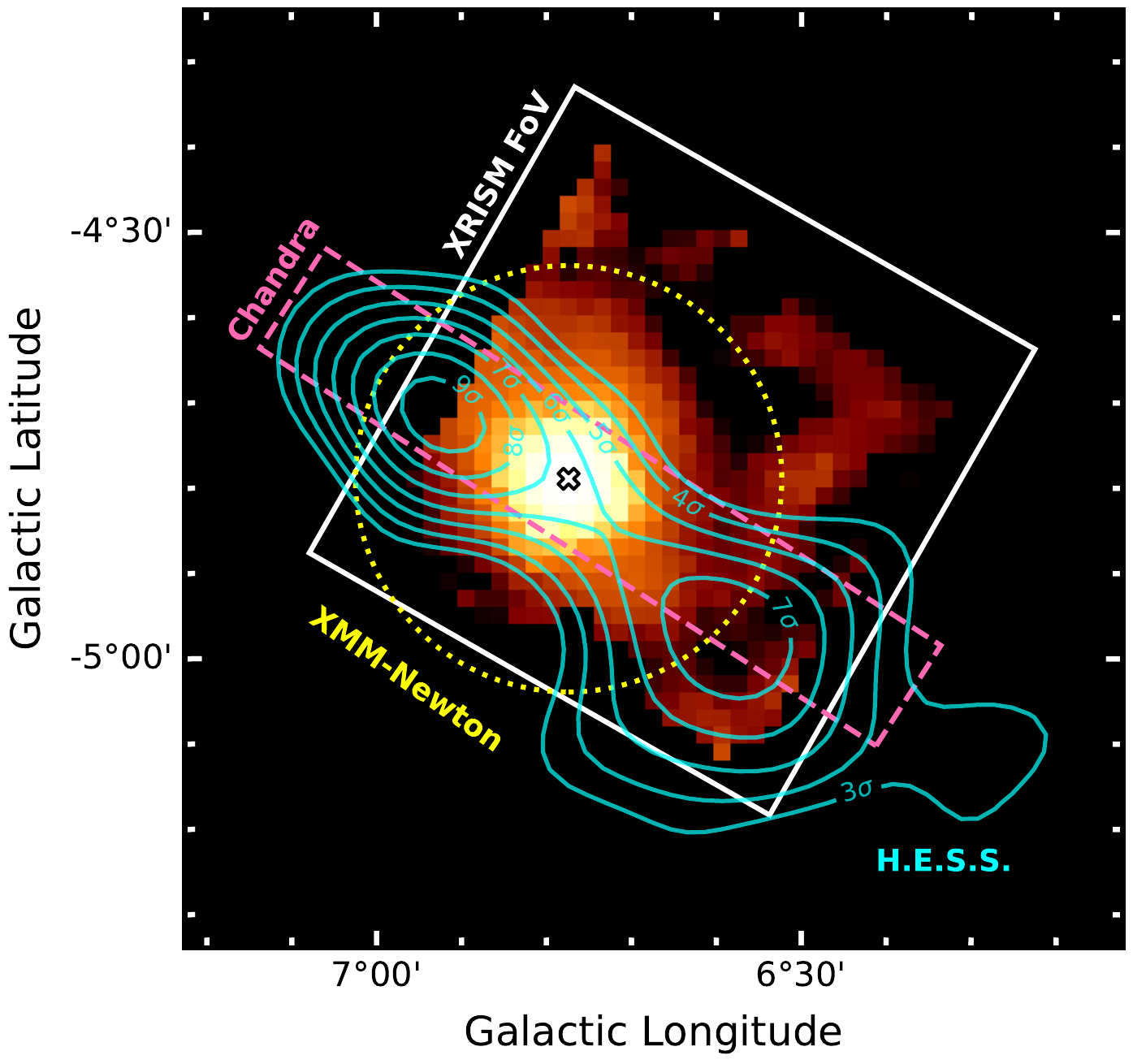}
	\caption{\label{fig:xray} \xray coverage.  The orange colours show the XRISM image after the particle-background subtraction and vignetting correction \cite[Figure~2c from~][]{Suzuki2025}. The FoV of XRISM is denoted with a white box. The blue contours are the same as shown in black in Figure~\ref{fig:significance}, outlining the \hess emission. Existing coverage by \textit{Chandra} and \textit{XMM-Newton} is shown with pink dashed and yellow dotted lines, respectively. }
\end{figure}

Figure~\ref{fig:xray} compares the XRISM image of the \vsgr region to the \hess contours. As can be seen, the morphology of the XRISM emission does not closely follow that of the gamma~rays, nor extends as much within the XRISM FoV. A dedicated analysis of the archival \textit{Chandra} data~\citep{Suzuki2025} revealed flux levels similar to those obtained with XRISM, although with higher systematic uncertainties due to the absence of background regions outside of the gamma-ray emission.

\vspace{-0.3 cm}
\section{Details on the estimation of the gas density} 
\label{app:gas}
\subsection{Molecular gas}
$^{12}$CO(J=1-0) measurements were obtained with the Nobeyama Radio Observatory (NRO) 45~m radio telescope between February and March 2025. An area of 0.8 $\times$ 1.4$\degree$ centred at R.A.=$18$h$19$m$28.8$s, Dec = $25$\degree$39$m$36$s was scanned in constant declination bands. The total observation time added up to roughly 20~h. The beamwidth of the NRO 45-m telescope is 15". For each pixel covered, a velocity with respect to the local standard of rest ($V_{\rm LSR}$) spectrum with resolution $\Delta V_{\rm LSR}$ = 0.2~km~s$^{-1}$ was obtained. The velocity coverage ranges from -150 to 150~km~s$^{-1}$.  The spatial and velocity grids have pixel sizes of 7.5" and 0.2~km~s$^{-1}$, respectively. Further details on the data-reduction steps follow those detailed in~\cite{Tsuji2025}. 

Assuming a distance of 6.2~kpc places \vsgr at a galacto-centric radius R$_{\mathrm{gal}}$ = 2.11~kpc (3.23 and 0.96~pc for distances of 5 and 8~kpc, respectively). At such small R$_{\mathrm{gal}}$, the Galactic rotation curve is relatively poorly constrained. We use the parametrisation provided by~\cite{Sofue2015} to transform the velocities in the measured spectra to distances because it is derived using more complete observational constraints in the R$_{\mathrm{gal}}<4$~kpc range. We will compare the results to those obtained when using different parametrisations~\citep{Brand1993,Reid2019} for completeness, but note that those are derived from few or no measurements at low R$_{\mathrm{gal}}$. The distance to \vsgr corresponds to $V_{\rm LSR}=64.6$~km~s$^{-1}$ (35 and 88~km~s$^{-1}$ for 5 and 8~kpc, respectively).

Figure~\ref{fig:gas_main} shows the velocity spectrum and a velocity-integrated map of $^{12}$CO around \vsgr. There is no significant signal in any range other than  (I) 4~km~s$^{-1}<V_{\rm LSR}<$10~km~s$^{-1}$, (II) 26~km~s$^{-1}<V_{\rm LSR}<$31~km~s$^{-1}$ and (III) -28~km~s$^{-1}<V_{\rm LSR}<$-24~km~s$^{-1}$. None of these ranges correspond to a distance consistent with any estimates of the distance to \vsgr regardless of the rotation curve used.

Velocity range (I) corresponds to a near and far distances of $1.7\pm0.62$ and $14.5\pm0.62$~kpc. The emission in that range is concentrated on a region of radius $\approx$0.1$\degree$ around the Galactic coordinates (6.393, -5.109)$\degree$, with a second, smaller bright spot around (6.233, -5.468)$\degree$. This newly discovered molecular cloud complex shows sub-structure and is likely made up of several smaller components.  
Velocity range (II) corresponds to near and far distances of $4.39\pm0.25$ and $11.80\pm0.25$~kpc, respectively. This velocity range is closest to, although still inconsistent with, the distance to \vsgr and the emission is concentrated on a small (<0.05$\degree$ radius) region around Galactic coordinates (6.369, -4.859)$\degree$. 
Velocity range (III) is just outside of the range of allowed velocities for that line of sight and all the rotation curves considered, which means the cloud has a non-negligible proper velocity. For relatively small values of the proper velocity, the closest allowed rotation velocity requires the cloud to be either very nearby ($<100$~pc) or very distant ($>15$~kpc). In any case, we can confidently state that this cloud does not explain the gamma-ray emission because the emission is again concentrated on a small (<0.04$\degree$ radius) region around Galactic coordinates (6.995, -4.638)$\degree$. The small size favours the large distance, to match the expected physical scales of molecular clouds of tens of pc~\citep{Heyer2015}, with 0.04$\degree$ corresponding to $\approx14$~pc at a distance of 20~kpc.
Due to the lack of correlation with the gamma-ray emission, and the inconsistent distances, a detailed study of these clouds is outside the scope of this paper and will be tackled by a follow-up paper~(Tsuji et al, in prep).

\begin{figure*}
	\centering
		\includegraphics[width=0.95\linewidth]{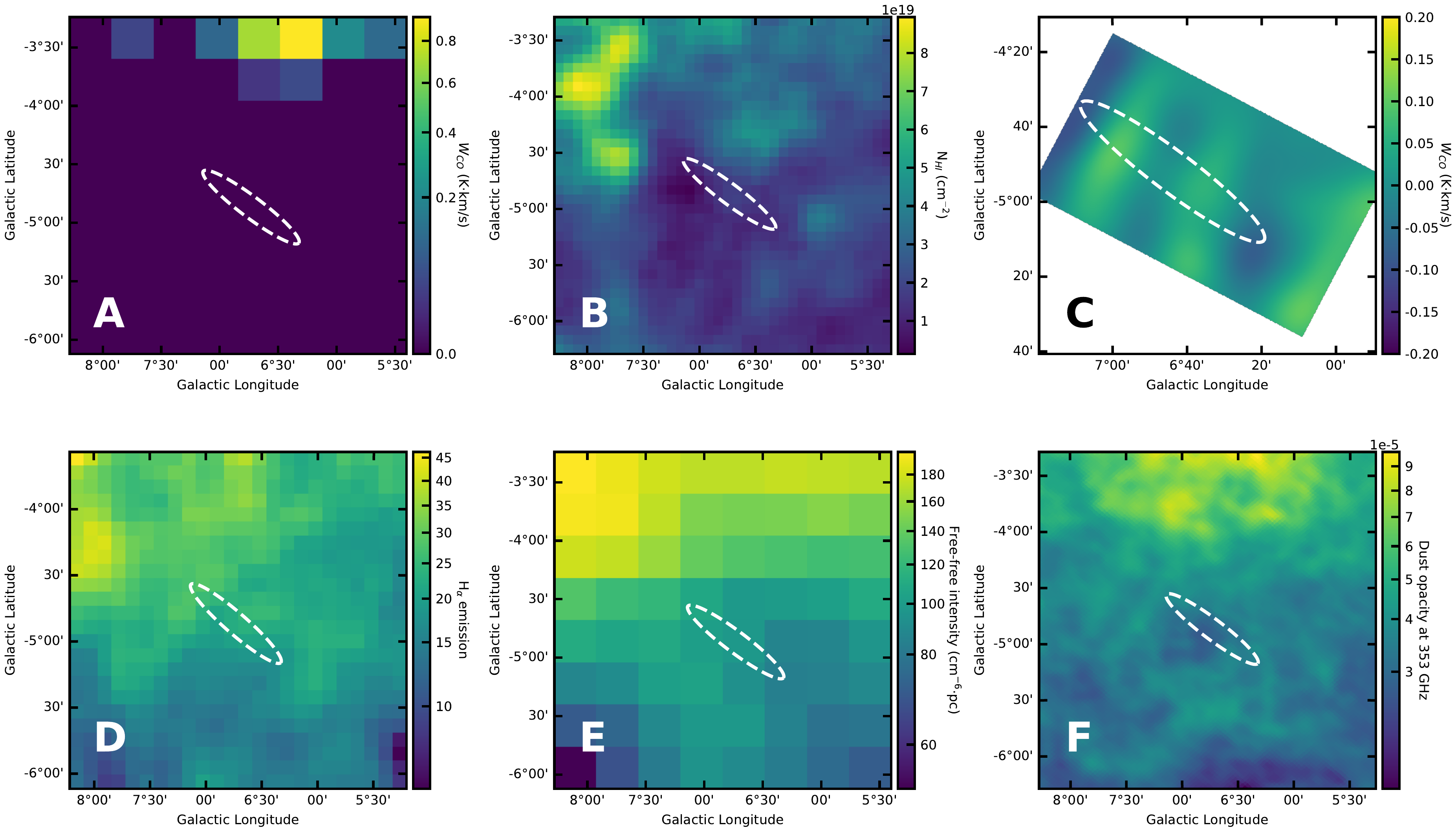}
	\caption{\label{fig:gas}  Gas around \vsgr. \textit{Panel A}: The velocity-integrated W$_{\rm CO}$ map from the DAME survey (all velocities). The dotted white line shows the 68\% containment region of model A. \textit{Panel B}: The same as panel A, but the map shows the hydrogen column density integrated in a range of 20~km~s$^{-1}$ around the velocity corresponding to the position of \vsgr. 
    \textit{Panel C}: The same as panel A, but the map shows the CO intensity (W$_{\rm CO}$) from the NRO data integrated in a range of 10~km~s$^{-1}$ around the velocity corresponding to the position of \vsgr. The map has been smoothed with a 0.1$\degree$ kernel to highlight the scales used to derive the density upper limit. The density upper limit was not derived using this map, but the noise level across all velocities.  
    \textit{Panel D}: The same as panel A, but the map shows the intensity of H$\alpha$ emission integrated along the line of sight from~\cite{Finkbeiner2003}. \textit{Panel E}: Same as panel A, but the map shows the intensity of free-free emission integrated along the line of sight from~\cite{PlanckCollaboration2016}. \textit{Panel F}: The same as panel A, but the map shows the dust opacity integrated along the line of sight from~\cite{PlanckCollaboration2016}.}
\end{figure*}

Our analysis of the dedicated NRO observations yielded no molecular cloud consistent with either the morphology of the gamma-ray emission or the distance to \vsgr. We can derive an upper limit to the average density in a region comparable to the gamma-ray emission from the noise level in the CO data. On scales comparable to the smallest dimension of the  gamma-ray emission (0.1$\degree$, 10.8~pc) the noise level across the data is  $\Delta T_{MB} =0.087$~K. We derive a 99\% containment upper limit on the CO intensity (W$_{CO}$) as W$_{CO, 99\%} < 3\times \Delta T_{MB} \times \sqrt{N_{ch}} \times \Delta V_{\rm LSR}$, where $N_{ch}$ is the number of velocity channels for integration (velocity range divided by resolution). We adopt a velocity range for the integration of 10~km~s$^{-1}$, larger than the typical spreads for molecular clouds~\citep{Heyer2015}, which results in W$_{CO, 99\%} < 0.3$~K~km~s$^{-1}$. Converting from CO to a column density of hydrogen requires an assumption for the  CO-to-H$_2$ conversion factor X$_{CO}$. In order to take into account fact that X$_{CO}$ depends on metallicity, which in turn depends on R$_{\mathrm{gal}}$, we adopt the functional form from~\cite{CO-conversion} with the same parameters as in~\cite{ctagps}. This results in a conversion factor X$_{CO}^{R_{\mathrm{gal}}=2.11~\mathrm{kpc}}= 3.5\times 10^{19}$ s~K$^{-1}$~km$^{-1}$~cm$^{-2}$ at the distance of \vsgr, consistent with Herschel estimates~\citep{Pineda2013} for low values of R$_{\mathrm{gal}}$. We assume that the depth of the cloud is the geometrical mean of the 68\% containment diameter along the major and minor axis of the gamma-ray emission, $l\approx 46$~pc, which yields n$_{\rm{H}}<0.18 \times l_{46}^{-1}$~cm$^{-3}$.

 For consistency, we also investigate the publicly available Dame Survey maps~\citep{Dame2022}. Since the angular resolution of these observations is significantly worse than that of NRO, small dim clouds might not be resolved, and in fact none of the clouds mentioned above are detected - there is no emission in a radius of 1$\degree$ around \vsgr (see Figure~\ref{fig:gas}, panel A). 

\subsection{HI, H$\alpha$, free-free emission and dust}
In addition to dedicated and archival CO observations, we consider the presence of atomic hydrogen through the study of 21-cm emission. We use the publicly available map from the HI4PI survey~\citep{Winkel2016} and follow the same procedure as for the CO maps, assuming optically thin gas. In the absence of lines in the relevant velocity range, we integrate the emission within a range of width 20~km~s$^{-1}$, motivated by the mean line-width found in a decomposition of the entire HI4PI survey data~\citep{Kalberla2018}. This is a conservative choice, as this value corresponds to the warm neutral medium (WNM), which is likely hotter than the environment around \vsgr~-- the mean line widths for the lukewarm and cold neutral media are lower by factors of more than 2 and 4, respectively~\citep{Kalberla2018}. Panel D in Figure~\ref{fig:gas} shows the value of N$_H$ integrated in a range of 20~km~s$^{-1}$ around the velocity corresponding to the position of \vsgr, where no neutral gas excess can be seen. We find n$_{\rm{HI}} \approx 0.08 \times l_{46}^{-1}$~cm$^{-3}$.

The HI column density map shown in panel B of Figure~\ref{fig:gas} and density estimate obtained differs significantly from those presented in~\cite{hawcv4641}. In that work, a velocity range of width 60~km~s$^{-1}$ was chosen for the integration, three times larger than the average line widths of the WNM~\citep{Kalberla2018}. While the velocity range corresponds to a physical distance of almost 1~kpc~\citep{Sofue2015}, a depth of only 20~pc was adopted, resulting in a higher density estimate ($\approx$1~cm$^{-3}$) than reported here. 

Furthermore, we check for the presence of ionised gas using the publicly available H$\alpha$ map from~\cite{Finkbeiner2003}. As can be seen in panel D of Figure~\ref{fig:gas}, we find no excess in the vicinity of \vsgr, indicating that, in addition to no neutral or molecular gas, there is also no ionised material.  There is no excess either in the Planck foreground maps\footnote{\url{https://wiki.cosmos.esa.int/planckpla2015/index.php/CMB_and_astrophysical_component_maps\#Astrophysical_foregrounds_from_parametric_component_separation}}\citep{PlanckCollaboration2016} of integrated free-free emission (Figure~\ref{fig:gas}, panel E)  or dust optical depth (Figure~\ref{fig:gas}, panel F), further confirming that \vsgr sits in a low-density environment.

\section{Summary of emission and acceleration scenarios considered.}
Given the large uncertainties and many scenarios discussed, we have collected some of the options in Table~\ref{tab:werner}, together with the main arguments against each scenario under a fully leptonic and fully hadronic assumption. This table is not an attempt to be exhaustive, with only simple scenarios taken into account.

\newpage
\begin{sidewaystable*}
 \caption[]{\label{tab:werner} Summary of emission and acceleration scenarios considered.}
 \resizebox{\textwidth}{!}{\begin{tabular}{ccccc|ll}
 \hline \hline
 \noalign{\vskip 1mm} 
\multicolumn{5}{c|}{ \textbf{Model}} & \multicolumn{2}{c}{ \textbf{Emission mechanism}}\\
 \hline 
\noalign{\smallskip}
  Jet orientation &  Jet speed & Jet length & Acceleration region &    Origin of morphology & issues if leptonic & issues if hadronic \\
 \hline 
 \hline 

\noalign{\smallskip}

  aligned with l.o.s. &  relativistic & $>$300~pc & near black hole &    advection in jet & energy in jet cocoon very large (Equation~\ref{eq:tjet}) & energy in jet cocoon very large \\
     &   & & &    & beaming should hide counter-jet (Table~\ref{tab:fit_params}) & beaming should hide counter-jet \\
     &   & & &    & requires weak magnetic field (B$\lesssim3\,\mu$G)  & energy in protons very large (but  \\
      &   & & &    &   &  benefits from beaming) \\
     &   & & &    &  & proton confinement \\
\hline
\noalign{\smallskip}
  aligned with l.o.s. &  relativistic & $<$300~pc & in lobes/cocoon &    filled lobes & energy in jet cocoon very large (Equation~\ref{eq:tjet}) & energy in filled cocoon very large \\
     &   & lobe size $\approx$100~pc& &    & requires weak magnetic field (B$\lesssim3\,\mu$G)  & energy in protons very large \\

     &   & & &    &  & proton confinement \\
\hline
\noalign{\smallskip}
  aligned with l.o.s. &  relativistic & any & near black hole &    escape from jets and diffusion  & unclear what creates special environment &  unclear what creates special environment\\
     &   & & or further out in jet &  into special environment  & unclear what determines source size &  unclear what determines source size\\

     &   & & &    & requires fine-tuning diffusion ($B\lesssim3\,\mu$G) & requires fine-tuning diffusion (B$\gtrsim100\,\mu$G) \\
          &   & & &    &  & energy in protons very large \\
     &   & & &    & & proton confinement \\

\hline
\noalign{\smallskip}  
perpendicular (hidden jet) &  at least mildly relativistic & (30-70)$(\sin{\phi})^{-1}$~pc & near black hole &   advection in jet & requires weak magnetic field (B$\lesssim3\,\mu$G) &  energy in protons very large  \\
     &   & & &    & radiative cooling at acceleration site  & proton confinement \\
 \hline

\noalign{\smallskip}  
perpendicular (hidden jet) &  at least mildly relativistic & (30-70)$(\sin{\phi})^{-1}$~pc & further out in jet &   slow advection within cocoon & requires weak magnetic field (B$\lesssim3\,\mu$G) &  energy in protons very large  \\
     &   & & &    & radiative cooling at acceleration site  & proton confinement \\
          &   & & &    & cooling faster than transport  &  \\

\hline
\noalign{\smallskip}  
perpendicular (hidden jet) &  at least mildly relativistic & $<$(30-70)$(\sin{\phi})^{-1}$~pc & further out in jet &   diffusion & requires weak magnetic field (B$\lesssim3\,\mu$G) &  energy in protons very large  \\
     &   & & &    &  & proton confinement \\

\noalign{\vskip 1mm} 
\hline

\end{tabular}}
\end{sidewaystable*}

\end{appendix}
\end{document}